\documentclass[11pt]{article}
\usepackage{amssymb}
\usepackage{bm}
\usepackage{algorithm}
\usepackage{algorithmic}

\usepackage{ascmac}
\usepackage{amsthm}
\usepackage{amsmath}
\usepackage{amssymb}
\usepackage[pdftex]{graphicx}

\newcommand{\vect}[1]{\hbox{\boldmath$#1$}}
\newcommand{\svect}[1]{\hbox{\scriptsize\boldmath$#1$}}
\newcommand{\vt}{\vect{t}}
\newcommand{\vs}{\vect{s}}

\newcommand{\vo}{\vect{\omega}}
\newcommand{\svt}{\svect{t}}

\newcommand{\svo}{\svect{\omega}}
\newcommand{\ZZ}{\mathbb{Z}}
\newcommand{\RR}{\mathbb{R}}
\newcommand{\NN}{\mathbb{N}}

\title{Directional Lifting Wavelet Transform for Image Edge Analysis}
\author{Kensuke Fujinoki\thanks{Department of Mathematical Sciences, Tokai University, 4-1-1, Kitakaname, Kanagawa, 259-1292 Japan} \and Keita Ashizawa\thanks{National Institute of Technology, Maizuru College, 234, Maizuru, Kyoto, 625-0016 Japan}}
\date{}

\begin{document}
\maketitle

\begin{abstract}
In this paper, we propose a new two-dimensional directional discrete wavelet transform that can decompose an image into 12 multiscale directional edge components.
The proposed transform is designed in a fully discrete setting and thus is easy to implement in actual computations.
The proposed transform is viewed as a category of redundant discrete wavelet transforms implemented by fast in-place computational algorithms by a lifting scheme that has been modified to incorporate redundancy.
The redundancy is limited to $(N \times J+1)/4$, where $N=12$ is the directional selectivity and $J$ is a decomposition level of the multiscale transform.
Numerical experiments in edge detection using various images demonstrate the advantages of the proposed method over some conventional standard methods.
The proposed method outperforms several conventional edge detection methods in identifying both the location and orientation of edges, and well captures the directional and geometrical features of images.

\end{abstract}

\section{Introduction}
Most of the data that we consider in computer vision is not completely random but has a certain correlated structure. 
One question that emerges in this research field is, ``How can such a correlation or feature of a digital object be extracted efficiently?''
Image analysis is a classical research field but still invites considerable attention because it is the basis for many related fields, including computer vision.
Local correlation structures in an image appear as sharp transitions or singularities, such as edges, and thus edges are important features that can be a clue revealing the finer details of the image itself.
Edges consist of directional components, such as lines and curves, which intrinsically contain geometrical features of two-dimensional data.

Many methods have been proposed to represent image edges in the field of signal processing \cite{dir_filbank,canny,steerable}.
In particular, wavelet-based methods have shown substantial success in multiscale image analysis including edge detection, compression, and denoising \cite{SMbook,filterbank3}.
However, wavelets do not provide good directional selectivity, which results in some failure in geometrical image analysis.
To alleviate this problem, several directional wavelet-based methods have been proposed, such as Mallat's multiscale edge detection method \cite{MallatEdge}, two-dimensional directional wavelets \cite{antoine}, and dual-tree complex wavelets \cite{dualtree}, which use a continuous wavelet transform or complex number approach.
Additional geometrical multiscale approaches have also been proposed, including curvelets \cite{curvelet1st,curvelet2,curvelet3}, contourlets \cite{contourlet}, bandlets \cite{bandlet}, and shearlets \cite{shearlet2,shearlet1}.
One of the key ideas behind these methods is to allow redundancy in their construction. 
This allows for more flexibility in the design of a wavelet transform, such as having good directional selectivity.
Although they frequently outperform the traditional discrete wavelet transform (DWT) in geometrical analysis, the redundancy makes a transform or a system computationally expensive.

In this paper, we focus primarily on an efficient wavelet-based edge analysis method using redundancy with less computational cost.
Preliminary results related to the present study have been published previously \cite{FujiAshi,FujiAshi2}.
We present a compatible redundant wavelet transform, which improves the directional selectivity and reduces the computational cost of the traditional DWT.
To do this, we consider the lifting scheme \cite{WS,WS2} in a redundant setting in two dimensions.
The lifting scheme allows for an efficient implementation of the DWT as well as a framework for custom design of the DWT, and is therefore used in the design of directional transforms \cite{dirlift2,FujiLift,FujiTri,dirlift1}.
By extending the properties of the lifting scheme, we show that we can achieve a good trade-off between the redundancy and directional selectivity of the DWT.
The proposed lifting-based efficient algorithms into the redundant wavelet transform  should result in further performance improvements in image edge analysis.

The remainder of this paper is organized as follows.
We begin with a brief review of the lifting scheme and the DWT in Section 2.
In Section 3, we consider the extension of the directional selectivity of the DWT implemented by the lifting scheme in two dimensions.
In Section 4, we apply the proposed method to image edge analysis and compare the performance with that of some conventional methods.
Various numerical experimental results are shown, and the advantages of the proposed method are discussed.
Finally, concluding remarks are given in Section 5.

\section{Lifting wavelet transform (LWT)}
Let $\{c_j[n]\}_{n\in\ZZ}$ be a signal with resolution level $j\in\NN\cup \{0\}$.
We define 
a $2\pi$ periodic function of a signal $\{c_j[n]\}_{n\in\ZZ}$ as
$$
C_j({\omega})
= \sum_{n\in\ZZ}\, c_j[n]\, e^{-i{\omega}n},\quad {\omega}\in\RR.
$$
The polyphase representation \cite{Polyphase} of a signal $\{c_j[n]\}_{n\in\ZZ}$
is given by 
\begin{equation}\label{polyphase1D.eq}
C_j\left({{\omega}}\right)
=
C_{e,j}\left(2{\omega}\right)
+
C_{o,j}\left(2{\omega}\right)
e^{-i{\omega}n},
\end{equation}
where
\begin{equation}
C_{e,j}(\omega) = \sum_{n\in\ZZ}c_j[2n]e^{-i\omega n},\quad
C_{o,j}(\omega) = \sum_{n\in\ZZ}c_j[2n+1]e^{-i\omega n}. 
\label{polydecom1D.eq}
\end{equation}
Here \eqref{polydecom1D.eq} is called the polyphase decomposition of a signal $\{c_j[n]\}_{n\in\ZZ}$, and 
components $C_{e, j}(\omega)$ and $C_{o, j}(\omega)$ are called the polyphase components of the signal.
 
The lifting wavelet transform (LWT) addresses the successful handling of polyphase components individually.
The LWT can be defined in both the time domain and the frequency domain, and is fully compatible with the both domains. 
In the frequency domain, the scheme of the LWT corresponds to the factorization of a polyphase matrix assembled by polyphase components of Laurent polynomials \cite{ID}.
To get a closer look at the structure of the LWT, the frequency domain is often used, but we work here in the time domain because it is more intuitive to explain the structure \cite{WS3}.

In the time domain, the LWT can be realized in four steps, as follows:
\begin{description}
\item[Step 1 (split):]\quad
$
c_{e,j-1}[n]=c_j[2n],\quad c_{o,j-1}[n]=c_j[2n+1].
$
\item[Step 2 (predict):]\quad
$
d_{j-1}[n]=c_{o,j-1}[n]-\left(\mathcal{P}\, c_{e,j-1}\right)[n].
$
\item[Step 3 (update):]\quad
$
c_{j-1}[n]=c_{e,j-1}[n]+\left(\mathcal{U}\, d_{j-1}\right)[n].
$
\item[Step 4 (scaling):]\quad
$
c_{j-1}[n]=K c_{j-1}[n], \quad d_{j-1}[n]=1 / K d_{j-1}[n],\quad K\in\RR.
$
\end{description}
Here, the two operators $\mathcal{P}$ and $\mathcal{U}$ are called lifting operators. In addition, $\mathcal{P}$ is 
a prediction operator, and $\mathcal{U}$ is an update operator.
The two sequences $\{c_{j-1}[n]\}_{n\in\ZZ}$ and $\{d_{j-1}[n]\}_{n\in\ZZ}$ are called the coarse and detail components, respectively.

The first step of the LWT splits a signal into an even-indexed component $\{c_{e,j-1}[n]\}_{n\in\ZZ}$ and an odd-indexed component $\{c_{o,j-1}[n]\}_{n\in\ZZ}$, which we refer to as split sequences.
The LWT deals with these two split sequences, $\{c_{e,j-1}[n]\}_{n\in\ZZ} $ and $\{c_{o,j-1}[n]\}_{n\in\ZZ}$, which correspond to polyphase components $C_{e,j}(\omega)$ and $C_{o,j}(\omega)$.
In Step 2, the odd-indexed sequence $\{c_{o,j-1}[n]\}_{n\in\ZZ}$ is predicted by the predictor $\mathcal{P}$ that operates on the even-indexed sequence $\{c_{o,j-1}[n]\}_{n\in\ZZ}$.
Then, in Step 3, the even-indexed sequence $\{c_{e,j-1}[n]\}_{n\in\ZZ}$ is updated by the updater $\mathcal{U}$, which operates on the odd-indexed sequence $\{d_{j-1}[n]\}_{n\in\ZZ}$ that has been predicted by the predictor $\mathcal{P}$.
Finally, the results are scaled for a certain normalization in Step 4.

The inverse lifting wavelet transform (ILWT) is defined by reversing these forward steps:
\begin{description}
\item[Step 1 (re-scaling):]\quad
$
c_{j-1}[n]=1/K c_{j-1}[n], \quad d_{j-1}[n]= K d_{j-1}[n].
$
\item[Step 2 (undo predict):]\quad
$
c_{e,j-1}[n]=c_{j-1}[n]-\left(\mathcal{U}\, d_{j-1}\right)[n].
$

\item[Step 3 (undo update):]\quad
$
c_{o,j-1}[n]=d_{j-1}[n]+\left(\mathcal{P}\, c_{e,j-1}\right)[n].
$

\item[Step 4 (merge):]\quad
$
c_j[2n]=c_{e,j-1}[n],\quad c_j[2n+1]=c_{o,j-1}[n].
$
\end{description}

One of the main advantages of the LWT is that this scheme is always invertible, which means that we can easily design the inverse transform that recovers the original signal.
By constructing appropriate lifting operators $\mathcal{P}$ and $\mathcal{U}$, the lifting scheme gives exactly the same results as those in the DWT implemented by Mallat's decomposition algorithms \cite{Mallat89}, which can be written in terms of a low-pass filter $\{h[n]\}_{n\in\ZZ}$ and a high-pass filter $\{g[n]\}_{n\in\ZZ}$ as
\begin{equation}\label{mallat.eq}
\begin{split}
c_{j-1}[n]&=\sum_{l \in \mathbb{Z}} \overline{h[l-2n]}\, c_{j}[l], \\
d_{j-1}[n]&=\sum_{l \in \mathbb{Z}} \overline{g[{l-2 n}]}\, c_{j}[l],
\end{split}
\end{equation}
where $\overline{h[l]}$ is the complex conjugate of $h[l]$.
With a dual low-pass filter $\{\tilde h[n]\}_{n\in\ZZ}$ and a dual high-pass filter $\{\tilde g[n]\}_{n\in\ZZ}$, the reconstruction is written as
$$
c_{j}[n]=\sum_{l \in \mathbb{Z}}\left(\tilde{h}[{n-2 l]}\, c_{j-1}[l]+\tilde{g}[{n-2 l}]\, d_{j-1}[l]\right).
$$
Note that, unlike the LWT, this implementation requires finding a pair of dual low-pass and high-pass filters $\{\tilde h[n],\tilde g[n]\}_{n\in\ZZ}$ that is an appropriate pair of primal filters $\{h[n],g[n]\}_{n\in\ZZ}$ in order to achieve an exact reconstruction.
That is why the LWT has a clear advantage, because, in general, the existence of the dual filters is not always guaranteed.

\section{Directional lifting wavelet transform}

Assume that a digital image $f$ is given on the plane $\ZZ^2$.
We denote $\{c_j[\bm{t}]\}_{\bm{t}\in\ZZ^2}$ by a sequence of pixel values of an image.
As in one dimension, we define 
a $2\pi$ periodic function of a signal $\{c_j[\vt]\}_{\bm{t}\in\ZZ^2}$ as
$$
C_j(\bm{\omega})
= \sum_{\bm{t}\in\ZZ^2}\, c_j[\vt]\, e^{-i\bm{\omega}\cdot\bm{t}},\quad \bm{\omega}\in\RR^2.
$$
The polyphase representation \eqref{polyphase1D.eq} for a signal $\{c_j[\vt]\}_{\bm{t}\in\ZZ^2}$
is expressed as 
\begin{equation}\label{polyphase.eq}
C_j\left({\bm{\omega}}\right)
=
\sum_{m=0}^3
C_{m,j}\left(2\bm{\omega}\right)
e^{-i\bm{\omega}\cdot\bm{t}_m},
\end{equation}
where 
$
\bm{t}_0=\bm{0},\,\bm{t}_1 = (1,\,0)^T,\,
\bm{t}_2 = (0,\,1)^T,\,
\vt_3\in\{
\vt_1+\vt_2,\,
-\vt_1+\vt_2,\,
\vt_1-\vt_2,\,
-\vt_1-\vt_2\}
$
and
\begin{equation}
C_{m, j}(\vo) = \sum_{\bm{t}\in\ZZ^2}c_j[2\vt+\vt_m] e^{-i\bm{\omega}\cdot\bm{t}},
\quad m=0,1,2,3
\label{polydecom.eq}
\end{equation}
are four polyphase components.
By incorporating the four polyphase components $\{C_{m, j}\}_{m=0,1,2,3}$ into the scheme of the LWT, we obtain the two-dimensional DWT.
For $m=0$, the function 
$C_{0, j}(\vo)$ is considered to be an even-indexed component, whereas 
$\{C_{m, j}\}_{m=1,2,3}$ are odd-indexed components.
Each element of the odd components $\{C_{m, j}\}_{m=1,2,3}$ is also considered to be a directional phase component along the direction $\{\vt_m\}_{m=1,2,3}$.
Thus, we say that the directional selectivity of the two-dimensional DWT is $N=3$.

\subsection{Redundant polyphase decomposition}
\begin{figure}[t]
\begin{center}
\includegraphics[width=7cm]{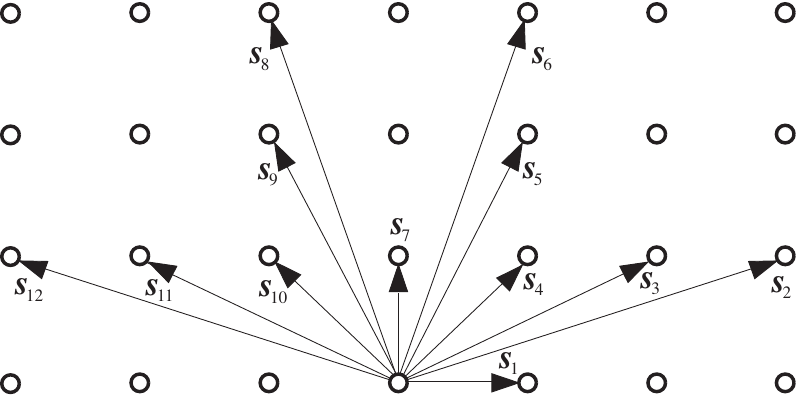}
\caption{
Arrangements of vectors $\{\vs_k\in\ZZ^2\}_{k\in D}$
}
\label{vector.fig}
\end{center}
\end{figure}

We consider extending the directional selectivity of the DWT to $N=12$.
We introduce directional vectors $\{\vs_{m}\in\ZZ^2\}_{0\leq m\leq N}$ defined by the linear combination of the vectors $\vt_1$ and $\vt_2$, so that 
each vector $\bm{s}_m$ approximately represents the direction
$$
\theta\approx\left(\frac{180\left(d-1\right)}{N}\right)^{\circ},\quad d\in D=\{\ell\in \ZZ \mid 1\leq 
\ell\leq N\},
$$
where $\theta$ is determined depending on the combination of the vectors $\bm{t}_1$ and $\bm{t}_2$, e.g., 
$\theta=\arctan\left(\|\bm{t}_2\|/\|\bm{t}_{1}\|\right) = 45^{\circ}$.
The arrangements of $\{\vs_{m}\}_{0\leq m\leq N}$ on a lattice $\ZZ^2$ are shown in Figure \ref{vector.fig}.

By using these vectors $\{\vs_{m}\}_{0\leq m\leq N}$,
we can rewrite the polyphase decomposition \eqref{polydecom.eq} as
\begin{equation*}
\widetilde C_{m, j}(\vo) = \sum_{\bm{t}\in\ZZ^2}c_j[2\vt+\vs_m] e^{-i\svo\cdot\svt},
\quad 0\leq m\leq N.
\label{polyphase2.eq}
\end{equation*}
Due to the periodicity of $e^{-i\svo\cdot\svt}$, these polyphase-like components $\{\widetilde C_{m, j}\}_{0\leq m\leq N}$ are no longer disjoint sets and are thus redundant.
As indicated by the polyphase representation \eqref{polyphase.eq}, the number of independent odd components in two dimensions is three. 
The redundant components $\{\widetilde C_{m,j}\}_{m\in D}$ are classified into three independent sets $\{\widetilde C_{m,j}\}_{m\in D_n},\,n=1,2,3$ with
\begin{align*}
D_{1}&=\{4 \ell+1 \mid 0 \leq \ell \leq N / 6\} ,\\ D_{2}&=\{4 \ell+3 \mid 0 \leq \ell \leq N / 6\}, \\ D_{3}&=\{2 \ell+2 \mid 0 \leq \ell \leq N / 2\}.
\end{align*}
Thus, the equality \eqref{polyphase.eq} does not hold.
However, instead of having redundancy, we have 12 directional components $\{\widetilde C_{m, j}\}_{m\in D}$.
This implies that the LWT using these polyphase-like components has some redundancy but has more directional selectivity.

\subsection{Redundant directional LWT}\label{DLWT.sec}
Here, we define the redundant directional LWT based on the redundant polyphase decomposition.
Since we have 12 odd-like components,
for a signal $\{x[\vt]\}_{\svt\in\ZZ^2}$, we define a set of 12 prediction operators
$\,\{\mathcal{P}_k\}_{k\in D}$ and 12 update operators $\{\mathcal{U}_k\}_{k\in D}$.
We consider linear bounded operators as prediction and update operators, which are two-dimensional discrete convolution operators defined by
\begin{align*}
\left( \mathcal{P}_k \,x \right) [\vt] &=\sum_{\bm{\ell}\in{\ZZ^2}}\,p_k[\bm{\ell}] \,x[\vt-\bm{\ell}],
\\
\left( \mathcal{U}_k \,x \right) [\vt] &= 
\sum_{\bm{\ell}\in{\ZZ^2}}\,u_k[\bm{\ell}] \,x[\vt-\bm{\ell}],
\end{align*}
where 
$\{p_k[\bm{t}]\in\RR\mid k\in D,\,\bm{t}\in{\ZZ^2}\}$ are referred to as prediction filters and 
$ \{u_k[\bm{t}]\in\RR\mid\,k\in D,\,\bm{t}\in{\ZZ^2}\}$ are referred to as update filters.
We assume that all of the filters have finite impulse responses.

The LWT modified for our situation can be described as follows:
\begin{description}
\item[Step 1 (split):] Decompose a signal $\{c_j[\vt]\}_{\svt\in{\ZZ^2}}$ into even component $\{c_{0,j-1}[\vt]\}_{\svt\in{\ZZ^2}}$ and $N$ directional odd components $\{c_{k,j-1}[\vt]\}_{\svt\in\ZZ^2,k\in D}$ by a split operator $\mathcal{S}_N$ defined by
\begin{equation}\label{split.eq}
\left( \mathcal{S}_N \,c_{j}\right)[\bm{t}]
=\{c_{0,j-1}[\vt], \,c_{k,j-1}[\vt]\}_{\bm{t}\in\ZZ^2,k\in D},
\end{equation}
where
$$
c_{k,j-1}[\vt]= c_j[2\vt+\vs_k], \quad 0\leq k \leq N.
$$

\item[Step 2 (predict):] Calculate the $N$ detail components $\{d_{k,j-1}[\vt]\}_{\svt\in{\ZZ^2},k\in D}$ using the prediction operators $\{\mathcal{P}_k\}_{k\in D}$:
\begin{equation}
d_{k,j-1}[\vt]=c_{k,j-1}[\vt]-\left(\mathcal{P}_k \, c_{0,j-1}\right)[\vt].
\label{predict2d.eq}
\end{equation}

\item[Step 3 (update):] Calculate the coarse component $\{c_{j-1}[\vt]\}_{\svt\in{\ZZ^2}}$ using the update operators $\{\mathcal{U}_k\}_{k\in D}$ that are applied to the results of the prediction in Step 2:
\begin{equation}
c_{j-1}[\vt] = c_{0,j-1}[\vt]
+\sum_{n=1}^3\left(\alpha_n\sum_{k\in D_n}\left(\mathcal{U}_k \,d_{k,j-1}\right)[\vt]\right),
\label{update2d.eq}
\end{equation}
where $\alpha_n,n=1,2,3$ are parameters that are adjusted so that each sum for the subsets $D_n$ gives its average.

\item[Step 4 (scaling):]  Apply scaling to the output of the prediction and update steps by $K\in\RR$ for normalization:
$$
c_{j-1}[\vt]=K c_{j-1}[\vt], \quad d_{k,j-1}[\vt]=1 / K d_{k,j-1}[\vt].
$$
\end{description}
These steps can be iterated to an arbitrary decomposition level $J\geq 1$.
As a result, the following sequences of coefficients are obtained: 
$$
c_{j}[\bm{t}]
\mapsto
\left\{d_{k, j-1}[\bm{t}],\,d_{k, j-2}[\bm{t}],\,\ldots,\,d_{k, j-J}[\bm{t}],\,c_{j-J}[\bm{t}]\right\}_{\bm{t}\in\ZZ^2,k \in D}.
$$
We refer to this transform as a directional lifting wavelet transform (DLWT).
The DLWT deals with overlapped multi-phase components
$\{c_{k,j-1}[\vt]\}_{\bm{t}\in\ZZ^2,0\leq k\leq N}$ decomposed by the split \eqref{split.eq} that are a redundant representation of a signal $\{c_{j}[\bm{t}]\}_{\bm{t}\in\ZZ^2}$.
The 12 odd-like components $\{c_{k,j-1}[\vt]\}_{\bm{t}\in\ZZ^2,k\in D}$ can be viewed as directional components, and each detail component $\{d_{k,j-1}[\vt]\}_{\svt\in{\ZZ^2},k\in D}$ obtained by the prediction \eqref{predict2d.eq} is expected to reveal the directional correlation of a signal.
The update step of \eqref{update2d.eq} will give a smooth coarse signal $\{c_{j-1}[\vt]\}_{\bm{t}\in\ZZ^2}$ calculated using the 12 directional detail components $\{d_{k,j-1}[\vt]\}_{\svt\in{\ZZ^2},k\in D}$, such that the average of a signal is maintained:
$
\sum_{\bm{t} \in \ZZ^2} c_{j-1}[\vt]=1/4 \sum_{\bm{t} \in \ZZ^2} c_{j}[\vt].
$

Since the DLWT is the redundant transform, the reconstruction is not unique.
There are several ways to reconstruct the original signal.
The inverse directional lifting wavelet transform (IDLWT) that reconstructs the original signal can be described as follows:

\begin{description}
\item[Step 1 (re-scaling):]
$
c_{j-1}[\vt]=1/K c_{j-1}[\vt], \quad d_{k,j-1}[\vt]=K d_{k,j-1}[\vt].
$

\item[Step 2 (undo update):]
$c_{0,j-1}[\vt]=c_{j-1}[\vt]
-\sum_{n=1}^3\left(\alpha_n\sum_{k\in D_n}\left(\mathcal{U}_k \,d_{k,j-1}\right)[\vt]\right).
$

\item[Step 3 (undo predict):]
$
c_{k,j-1}[\vt]=d_{k,j-1}[\vt]+\left(\mathcal{P}_k \, c_{0,j-1}\right)[\vt].
$

\item[Step 4 (merge):]
$
 \mathcal{S}_N^{-1}\,
\{c_{0,j-1}[\vt], \,d_{k,j-1}[\vt]\}_{\bm{t}\in\ZZ^2,k\in D}
=c_{j}[\bm{t}].
$

\end{description}

The DLWT has several advantages over the original LWT.
Obviously, the invertible structure of the scheme holds.
Here, we consider the discrete convolution operators for the predictors $\{\mathcal{P}_k\}_{k\in D}$ and updaters $\{\mathcal{U}_k\}_{k\in D}$, although no matter how these operators are chosen, the inverse transform for an exact signal reconstruction is guaranteed.
This means that the use of nonlinear operators is also possible.

Another advantage is that the DLWT has a low computational cost. 
The LWT can be implemented by the in-place algorithm of the lifting scheme, which does not require extra memory for computations because all calculations are performed by overwriting inputs with outputs. 
Suppose that 
${\bf even }=\{c_{0,j-1}[\vt]\}_{\bm{t}\in\ZZ^2}$
and 
${\bf odd }_k = \{c_{k,j-1}[\vt]\}_{\bm{t}\in\ZZ^2,k\in D}$ are two-dimensional arrays obtained by the split operation.
Then, written in terms of a programming language, the prediction and update steps of the DLWT use only ${\bf even }$ and ${\bf odd }_k$, as follows:
\begin{align*}
\bf  { odd_k }\,\, \texttt{-=}\,\,\,\mathcal{P}_k(\bf  { even });\quad
\bf  { even }\,\, \texttt{+=}\,\,\,\mathcal{U}_k(\bf  { odd }_k);
\end{align*}
Similarly, for the IDLWT, we have:
\begin{align*}
\bf  { even }\,\, \texttt{-=}\,\,\,\mathcal{U}_k(\bf  { odd }_k);\quad
\bf  { odd_k }\,\, \texttt{+=}\,\,\,\mathcal{P}_k(\bf  { even });
\end{align*}

As mentioned earlier, the DLWT is the redundant transform, but thanks to the use of the in-place algorithm, efficient implementation is possible.
The fast implementation algorithms for the DLWT and IDLWT are shown in Algorithm \ref{alg1} and Algorithm \ref{alg2}, respectively.
Moreover, the split \eqref{split.eq} gives the redundant representation of a signal but uses downsampling by a factor of 4, which provides a reasonable trade-off between the redundancy and the directional selectivity.
The redundancy of the DLWT is $(N\times J+1)/4$, which is better than that for the standard redundant DWT, i.e., the stationary wavelet transform, the redundancy of which is $3J+1$.
\begin{algorithm}[t]                      
\caption{DLWT}         
\label{alg1}       
\begin{algorithmic}[1]     
\renewcommand\algorithmicrequire{\textbf{Input:}}
\REQUIRE $\{c_j[\vt]\}_{\bm{t}\in\ZZ^2}$: signal, $L$: highest resolution level, $J$: number of decompositions.
\renewcommand\algorithmicrequire{\textbf{Require:}}
\REQUIRE $L\geq J,\,J\geq1$   
\FOR{$j\leftarrow L$ to $L-J+1$}
\STATE  $c_{j-1}^{(0)}[\vt] \leftarrow c_j[2\vt]$
\STATE  $d_{k,j-1}^{(0)}[\vt] \leftarrow c_j[2\vt+\bm{s}_k],\quad k\in D$
\FOR{$i \leftarrow 1$ to $n$}
\vspace{0.2cm}
\STATE $\displaystyle{d_{k,j-1}^{\,(i)}[\vt] \leftarrow d_{k,j-1}^{\,(i-1)}[\vt]-\sum_{\bm{v}\in\ZZ^2}\, p_k^{\left(i\right)}[\bm{v}]\,c_{j-1}^{\,(i-1)}[\vt-\bm{v}]},\quad k\in D$
\STATE $ \displaystyle{c_{j-1}^{\,(i)}[\vt] \leftarrow c_{j-1}^{\,(i-1)}[\vt]+\sum_{n=1}^3\left(\alpha_n\sum_{k\in D_n}\sum_{\bm{v}\in\ZZ^2}\, u_k^{\left(i\right)}[\bm{v}]\,d_{k,j-1}^{\,(i)}[\vt-\bm{v}]\right)}$
\ENDFOR
\STATE $c_{j-1}[\vt] \leftarrow K\,c_{j-1}^{\,(n)}\,[\vt]$
\STATE $d_{k,j-1}[\vt] \leftarrow 1/K \ d_{k,j-1}^{\,(n)}\,[\vt],\quad k\in D$
\ENDFOR
\end{algorithmic}
\end{algorithm}
\begin{algorithm}[t]                      
\caption{IDLWT}         
\label{alg2}     
\begin{algorithmic}[1]     
\FOR{$j= L-J+1$ to $L$}          
\STATE $c_{j-1}^{\,(n)}\,[\vt] \leftarrow 1/K\, c_{j-1}[\vt]$
\STATE $d_{k,j-1}^{\,(n)}\,[\vt] \leftarrow K \, d_{k,j-1}[\vt],\quad k\in D$
\FOR{$i \leftarrow n$ to $1$}
\vspace{0.2cm}
\STATE $ \displaystyle{c_{j-1}^{\,(i-1)}[\vt] \leftarrow c_{j-1}^{(i)}[\vt]-\sum_{n=1}^3\left(\alpha_n\sum_{k\in D_n}\sum_{\bm{v}\in\ZZ^2}\,u_k^{(i)}[\bm{v}]\,d_{j-1}^{\,(i)}[\vt-\bm{v}]\right)}$
\STATE $\displaystyle{
 d_{k,j-1}^{\,(i-1)}[\vt]\leftarrow d_{k,j-1}^{\,(i)}[\vt]+\sum_{\bm{v}\in\ZZ^2}\, p_k^{\left(i\right)}[\bm{v}]\,c_{j-1}^{\,(i-1)}[\vt-\bm{v}]},\quad k\in D$
\ENDFOR
\STATE$c_j[2\vt] \leftarrow c_{j-1}^{(0)}[\vt]$
\STATE$c_j[2\vt+\vt_1] \leftarrow $ select $d_{k,j-1}^{(0)}[\vt]$ with $k\in D_1$
\STATE$c_j[2\vt+\vt_2] \leftarrow $ select $d_{k,j-1}^{(0)}[\vt]$ with $k\in D_2$
\STATE$c_j[2\vt+\vt_3] \leftarrow $ select $d_{k,j-1}^{(0)}[\vt]$ with $k\in D_3$
\ENDFOR
\end{algorithmic}
\end{algorithm}

\subsection{Filter design methods}\label{ex1.sec}
Owing to the invertible structure of the DLWT, the present design method has flexibility and a high degree of freedom. 
In particular, since we have a set of 12 lifting operators $\{\mathcal{P}_k,\mathcal{U}_k\}_{k\in D}$ and their associated filters $\{p_k[\vt],u_k[\vt]\}_{\bm{t}\in\ZZ^2,k\in D}$, there are two options for making these operators nearly isotropic or anisotropic.
We focus on the isotropic method because we consider applying the proposed method to edge detection.
We attempt to design lifting filters with the same characteristics for each direction $\{\bm{s}_k\}_{k\in D}$.

To design prediction filters, we use the interpolating prediction introduced in \cite{WS}.
Let $M,\widetilde M=2n\,(n=0,1,2,\ldots)$, where $\widetilde M \leq M$, are even numbers.
The $M$-th-order interpolating prediction allows us to design a prediction filter of any order in the sense of the $M$-th-order Lagrange polynomials.
The filter has one-dimensional coefficients.
Now, we extend the filter to our directional settings by setting the one-dimensional coefficients of the interpolating prediction filter into directions $\{\bm{s}_k\}_{k\in D}$ on the lattice $\ZZ^2$.
The two-dimensional nonzero coefficients of the interpolating prediction filters $\{p_{k,M}[\vt]\}_{\bm{t}\in\ZZ^2,k\in D}$ for some $M$ are:
$$
\begin{cases}
p_{k,0}[\bm{0}]=1,&M=0,\\
p_{k,2}[\bm{0}]=\frac12, \quad p_{k,2}[\vs_k]=\frac12,&M=2,\\
p_{k,4}[-\vs_k]=-\frac{1}{16},\quad p_{k,4}[\bm{0}]=\frac{9}{16},\quad p_{k,4}[\vs_k]=\frac{9}{16},\quad p_{k,4}[2\vs_k]=-\frac{1}{16},&M=4.
\end{cases}
$$
The prediction \eqref{predict2d.eq} with these filters predicts directional odd components $\{c_{k,j-1}[\bm{t}]\}_{\bm{t}\in\ZZ^2,k\in D}$ by the $M$-th-order Lagrange interpolation using the even component $\{c_{0,j-1}[\vt]\}_{\bm{t}\in\ZZ^2}$.

The interpolating prediction allows us to construct update filters that are also based on the prediction filters.
The $\widetilde M$-th-order update filter for the interpolating prediction is defined by
\begin{equation*}
u_{k,\widetilde{M}}\,[\vt]=
p_{k,\widetilde{M}}\,[-\vt] / 4.
\end{equation*}
Updating $\eqref{update2d.eq}$ with these filters applied to  directional odd components $\{c_{k,j-1}[\vt]\}_{\bm{t}\in\ZZ^2,k\in D}$ provides a smooth even component $\{c_{j-1}[\vt]\}_{\bm{t}\in\ZZ^2}$ in the sense of the Lagrange interpolation.

\subsection{Examples}\label{ex2.sec}

Let us give some examples of the lifting filters with the interpolating prediction.
In the simplest example, which is the case of $M=\widetilde M=0$, 
the lifting operators $\{\mathcal{P}_{k},\,\mathcal{U}_{k} \}_{k\in D}$ become simple forms of 
\begin{align*}
\left(\mathcal{P}_{k} \,c_{0, j-1}\right)[\boldsymbol{t}]&=c_{0,j-1}[\bm{t}],\\ \quad\left(\mathcal{U}_{k} \,d_{k, j-1}\right)[\boldsymbol{t}]&=d_{k, j-1}[\vt]
/4.
\end{align*}
Thus, the prediction step is written as
$$
d_{k,j-1}[\vt]=c_{k,j-1}[\vt]-c_{0,j-1}[\vt],
$$
which simply calculates the difference between the even component and 12 directional odd components.
In other words, the prediction step predicts corrections of zeroth-order polynomials of a signal for each direction $\{\bm{s}_k\}_{k\in D}$. We refer to this step as a constant prediction.
The update step becomes
\begin{equation*}
c_{j-1}[\vt] = c_{0,j-1}[\vt]
+\sum_{n=1}^3\left(\alpha_n\sum_{k\in D_n}\,d_{k,j-1}[\vt]/4\right),
\end{equation*}
where $\alpha_1=\alpha_2=1/3$ and $\alpha_3=1/6$.
Note that we can always set these values for $\alpha_n$ for different orders of the interpolating prediction.

The next example is the case of $M=\widetilde M=2$.
We have 
\begin{align*}
\left(\mathcal{P}_{k} \,c_{0, j-1}\right)[\boldsymbol{t}]&=\frac{c_{0,j-1}[\bm{t}]+c_{0,j-1}[\bm{t}+\bm{s}_k]}{2},\\ \left(\mathcal{U}_{k} \,d_{k, j-1}\right)[\boldsymbol{t}]&=\frac{d_{k, j-1}[\vt-\bm{s}_k]+d_{k, j-1}[\vt]}{
8},
\end{align*}
and the DLWT is realized as  
\begin{align*}
d_{k,j-1}[\vt]&=c_{k,j-1}[\vt]-\frac{c_{0,j-1}[\vt]+c_{0,j-1}[\vt+\bm{s}_k]}{2},\\
c_{j-1}[\vt] &= c_{0,j-1}[\vt]
+\sum_{n=1}^3\left(\alpha_n\sum_{k\in D_n}\,\frac{d_{k,j-1}[\vt-\bm{s}_k]+d_{k,j-1}[\vt]}{8}\right).
\end{align*}
This is a linear prediction because directional odd components $\{c_{k,j-1}[\vt]\}_{\bm{t}\in\ZZ^2,k\in D}$ are predicted by a linear interpolation of two neighboring even components, $\{c_{0,j-1}[\vt]\}_{\bm{t}\in\ZZ^2}$ and $\{c_{0,j-1}[\vt+\bm{s}_k]\}_{\bm{t}\in\ZZ^2}$.
In this way, we can build a DLWT that has a different behavior by means of polynomial interpolation.

\subsection{Update-first construction}
Before closing this section, we mention one more option for the construction of the DLWT, which is an alternative method to build lifting filters for the DLWT.
In the filter design method described earlier, the prediction filters $\{p_k[\vt]\}_{\bm{t}\in\ZZ^2,k\in D}$ have two-dimensional filter coefficients, but prediction with these filters can essentially analyze only correlations along lines for $\{\bm{s}_k\}_{k\in D}$.
Here, we introduce a method by which to construct lifting filters $\{p_k[\vt],u_k[\vt]\}_{\bm{t}\in\ZZ^2,k\in D}$ that can analyze intrinsically local two-dimensional correlations of a signal.
The idea is to change the order of prediction and update steps, which is called the update-first form of the lifting scheme \cite{NLL}.

For simplicity, we introduce the construction using an update filter $\{u_k[\vt]\}_{\bm{t}\in\ZZ^2}$.
Let us first describe the construction of the update filter.
A two-dimensional update filter $\{u[\vt]\}_{\svt\in\ZZ^2}$ is constructed in two stages and is based on the B-spline function.
We first define a one-dimensional B-spline wavelet filter of order $r\in\NN$ in the Fourier domain as
$$
U_{0}(\omega) = 
\,e^{-{i \epsilon \omega/2} } \left(\cos \frac{\omega }{2}\right)^{r},
$$
where
$$
\left\{\begin{array}{ll}\epsilon=0, & \text { if } r \text { is even}, \\ \epsilon=1, & \text { if } r \text { is odd}.\end{array}\right.
$$
The coefficient form of this filter is given by 
$$
u_{0}[n] = \left(\mathcal{F}^{-1}\,
U_{0}(\omega)
\right)\left[n\right],
$$
where $\mathcal{F}^{-1}$ is the inverse Fourier transform of a $2\pi$ periodic function $X(\omega)$ defined as
$$
\left(\mathcal{F}^{-1}\,X\right)[n]=\frac{1}{2\pi}\int_{-\pi}^{\pi}X(\omega)\,e^{i\omega n}d{\omega}.
$$
We then obtain a B-spline-based two-dimensional update filter $\{u[\vt]\}_{\svt\in\ZZ^2}$ using the tensor product:
\begin{equation*}
u[\bm{t}]=\bm{u}_0\otimes\bm{u}_0,\quad \vt\in\ZZ^2.
\label{tensor.eq}
\end{equation*}

The update-first form of the DLWT using this filter is implemented in the following manner.
The coarse component $\{c_{j-1}[\vt]\}_{\svt\in{\ZZ^2}}$ of a signal is first computed using the convolution of the two-dimensional B-spline filter $\{u[\vt]\}_{\bm{t}\in\ZZ^2}$:
\begin{equation*}
c_{j-1}[\vt] =
\sum_{\bm{\ell} \in \ZZ^2} u[\bm{\ell}] c_{j}[2 \vt-\bm{\ell}].
\end{equation*}
Then, $N$ detail components $\{d_{k,j-1}[\vt]\}_{\svt\in{\ZZ^2},k\in D}$ are computed using the result of the update:
$$
d_{k,j-1}[\vt]=c_{k,j-1}[\vt]-\sum_{\bm{\ell}\in\ZZ^2}\,p_k[\bm{\ell}] \,c_{j-1}[\vt-\bm{\ell}].
$$
The prediction filters $\{p_k[\vt]\}_{\bm{t}\in\ZZ^2,k\in D}$ are constructed to have directionality along the direction $\{\bm{s}_k\}_{k\in D}$, as in the normal predict-first form of the DLWT, which is introduced in Section \ref{DLWT.sec}.
The interpolating prediction filters for the update-first form can be obtained as in Sections \ref{ex1.sec} and \ref{ex2.sec}.
In this setting, the parameters that we control are $r$ for the update filter, and $M$ for the interpolating prediction filters, which means that the update filter $\{u[\vt]\}_{\bm{t}\in{\ZZ^2}}$ is based on the $r$-th-order B-spline, while the prediction filters $\{p_k[\vt]\}_{\bm{t}\in{\ZZ^2},k\in D}$ are based on the $M$-th-order Lagrange polynomial.
These orders can be determined depending on an application.

\begin{figure*}[t]
\begin{center}
\includegraphics[width=12cm]{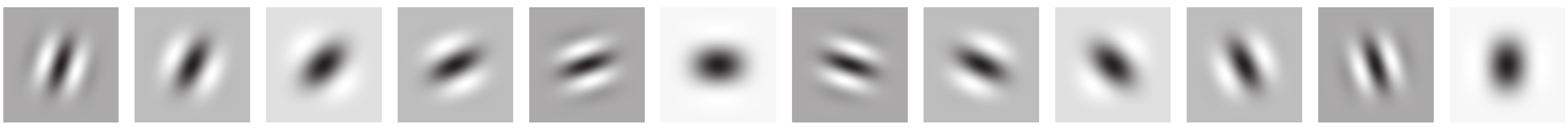}
\end{center}
\caption{
Frequency responses of high-pass filters $\{g_k[\bm{t}]\}_{\bm{t}\in\ZZ^2,k\in D}$ with $r=M=2$ in the update-first construction. 
}
\label{fres.fig}
\end{figure*}

Note that by the update-first approach for each detail component $\{d_{k,j-1}[\vt]\}_{\bm{t}\in\ZZ^2,k\in D}$ calculated by each prediction
with the updated coarse component $\{c_{j-1}[\vt]\}_{\bm{t}\in\ZZ^2}$
reveals the correlations of a two-dimensional signal along each direction $\{\bm{s}_k\}_{k\in D}$, even if
the prediction filters have essentially one-dimensional coefficients.
In other words, the results of the DLWT implemented by the update-first scheme is equivalent to those of the DWT by Mallat's decomposition algorithm \label{mallat.eq} using a low-pass filter $\{h[\bm{t}]\}_{\bm{t}\in\ZZ^2}$ and high-pass filters $\{g_k[\bm{t}]\}_{\bm{t}\in\ZZ^2,k\in D}$, both of which have two-dimensional coefficients, defined as  
\begin{equation*}
\begin{split}
c_{j-1}[\vt]&=\sum_{\bm{v} \in \mathbb{Z}^2} \overline{h[\bm{v}-2\vt]}\, c_{j}[\bm{v}], \\
d_{k,j-1}[\vt]&=\sum_{\bm{v} \in \mathbb{Z}^2} \overline{g_k[{\bm{v}-2 \vt}]}\, c_{j}[\bm{v}].
\end{split}
\end{equation*}
Figure \ref{fres.fig} shows the frequency response of the high-pass filters $\{g_k[\bm{t}]\}_{\bm{t}\in\ZZ^2,k\in D}$ with an order $r=M=2$ in the update-first construction. 
As expected, we see that they have a two-dimensional support with directional characteristics.

\section{Numerical experiments}
In this section, we present the results of numerical experiments applying the proposed method to edge detection in image processing.
The image size used in the experiments is $ 512\times 512$ pixels. Hence, the highest resolution level is set to $L=j=9$.
For simplicity, we do not use update-first construction for the DLWT.
\begin{figure*}[t]
\begin{center}
\includegraphics[width=12cm]{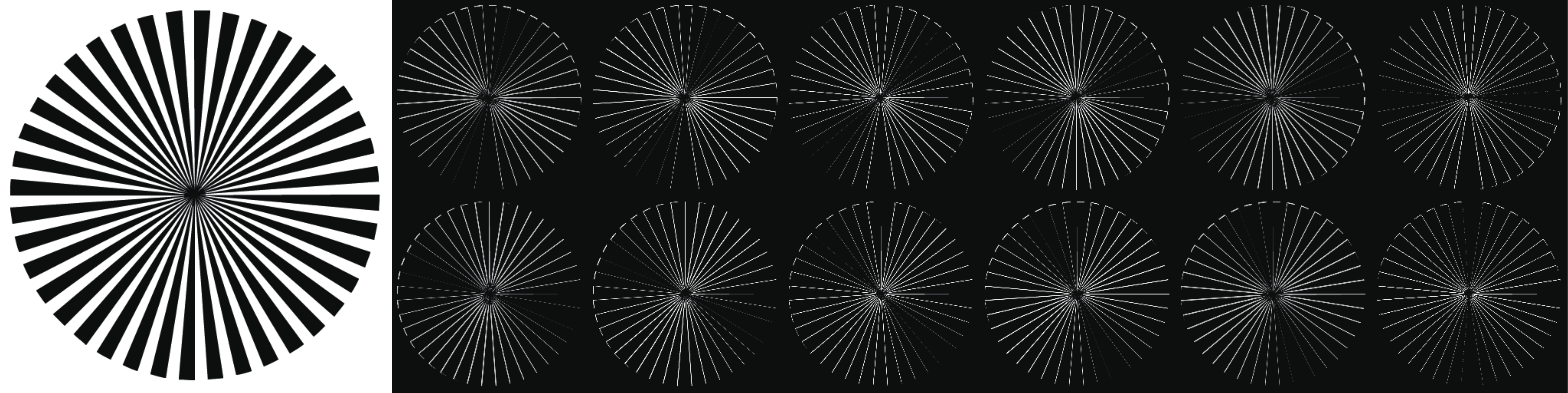}
\end{center}
\caption{
Original image $c_{j}[\vt]$ and its detail components $\{d_{k,j-1}[\vt]\}_{k\in D}$.
Top:
$k=6,\dots,1$, bottom: $k=12,\dots,7$.
}
\label{decom.fig}
\end{figure*}

\subsection{Image decomposition}

First, simple image decomposition is used to check that the proposed method is designed as intended. That is, we need to make sure that the image can be decomposed into 12 directions along the directional vectors $\{\bm{s}_k\}_{k\in D}$.
This can be checked by observing the correlation of the 12 detail components at a certain decomposition level $j$.
Figure \ref{decom.fig} shows the results for $\{d_{k,j-1}[\bm{t}]\}_{\bm{t}\in\ZZ^2,k\in D}$ obtained by the one-level DLWT to a radial circle image.
Obviously, each of the 12 detail components $\{d_{k,j-1}[\bm{t}]\}_{\bm{t}\in\ZZ^2,k\in D}$ corresponds to the directional component along each 12 directional vector $\{\bm{s}_k\}_{k\in D}$.
Thus, the DLWT can analyze the correlation of the image into 12 directional components, which suggests that this method can be a very effective for edge analysis of images.

\begin{figure*}[t]
\scriptsize
\begin{center}
\begin{tabular}{p{2.0cm}p{2.0cm}p{2.0cm}p{2.0cm}p{2.0cm}p{2.0cm}}
Original &$J=1$& $J=2$& $J=3$& $J=4$\\
\includegraphics[width=2.3cm]{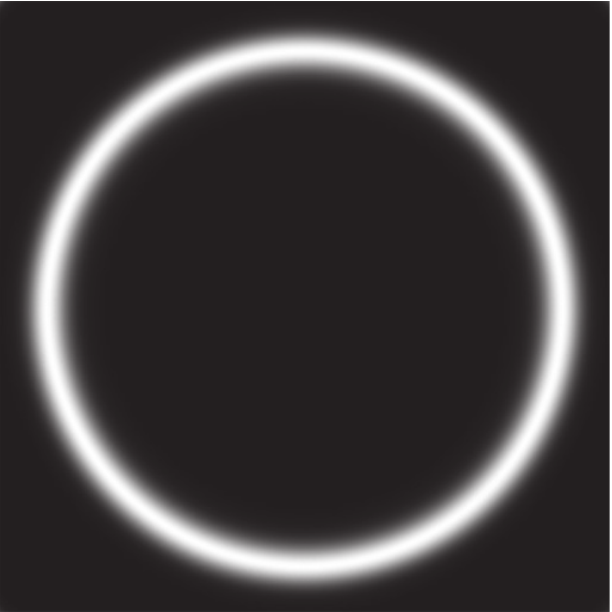}&
\includegraphics[width=2.3cm]{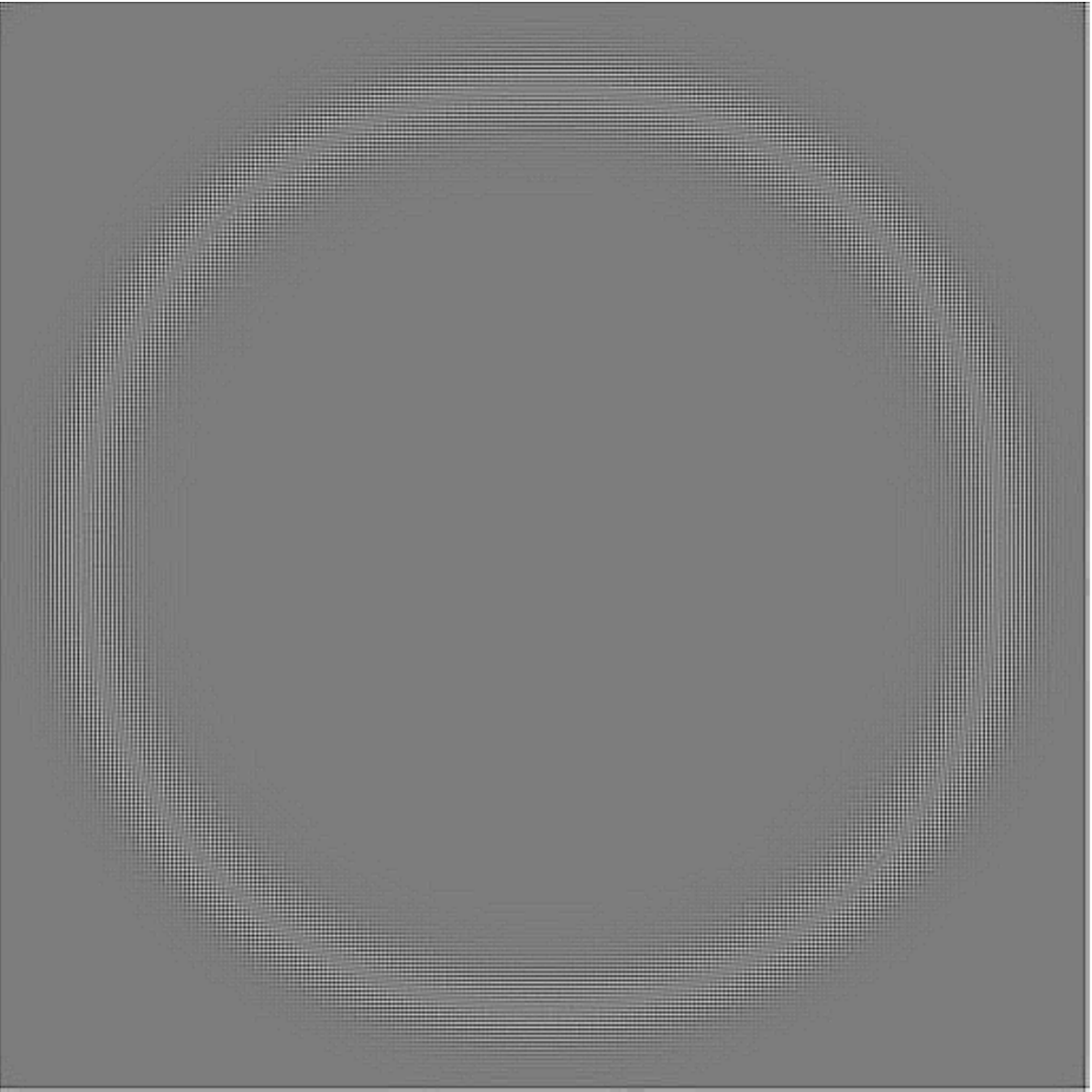}&
\includegraphics[width=2.3cm]{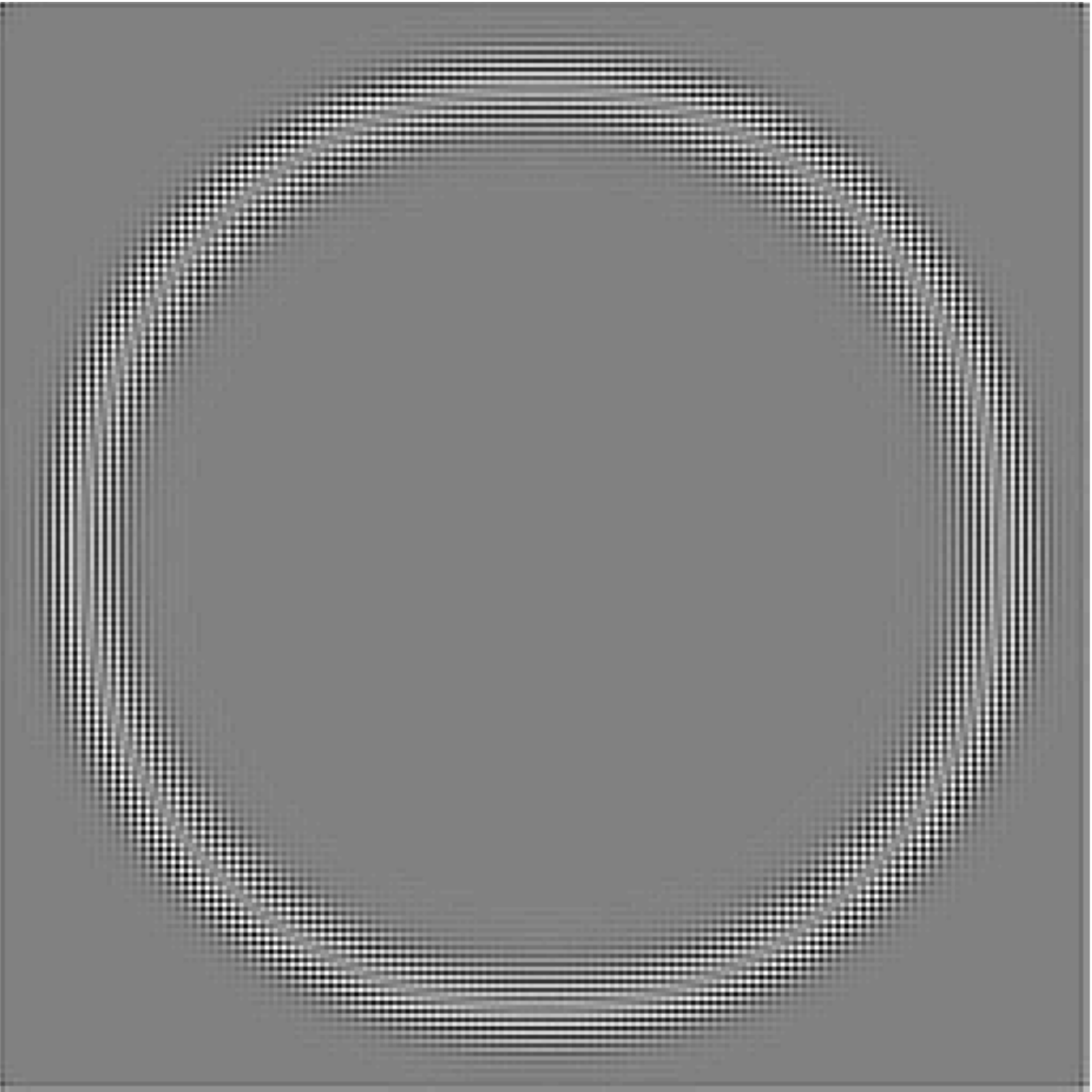}&
\includegraphics[width=2.3cm]{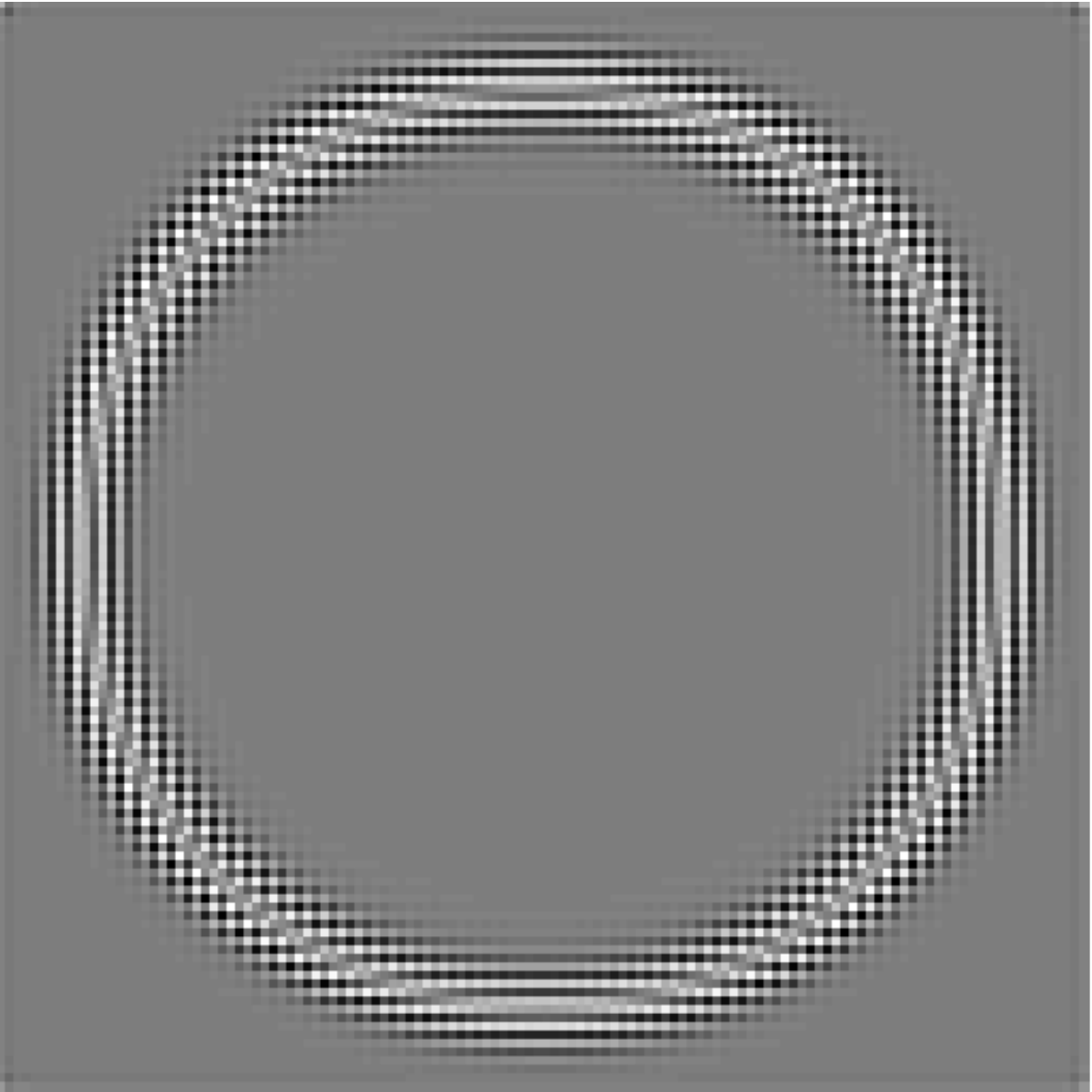}&
\includegraphics[width=2.3cm]{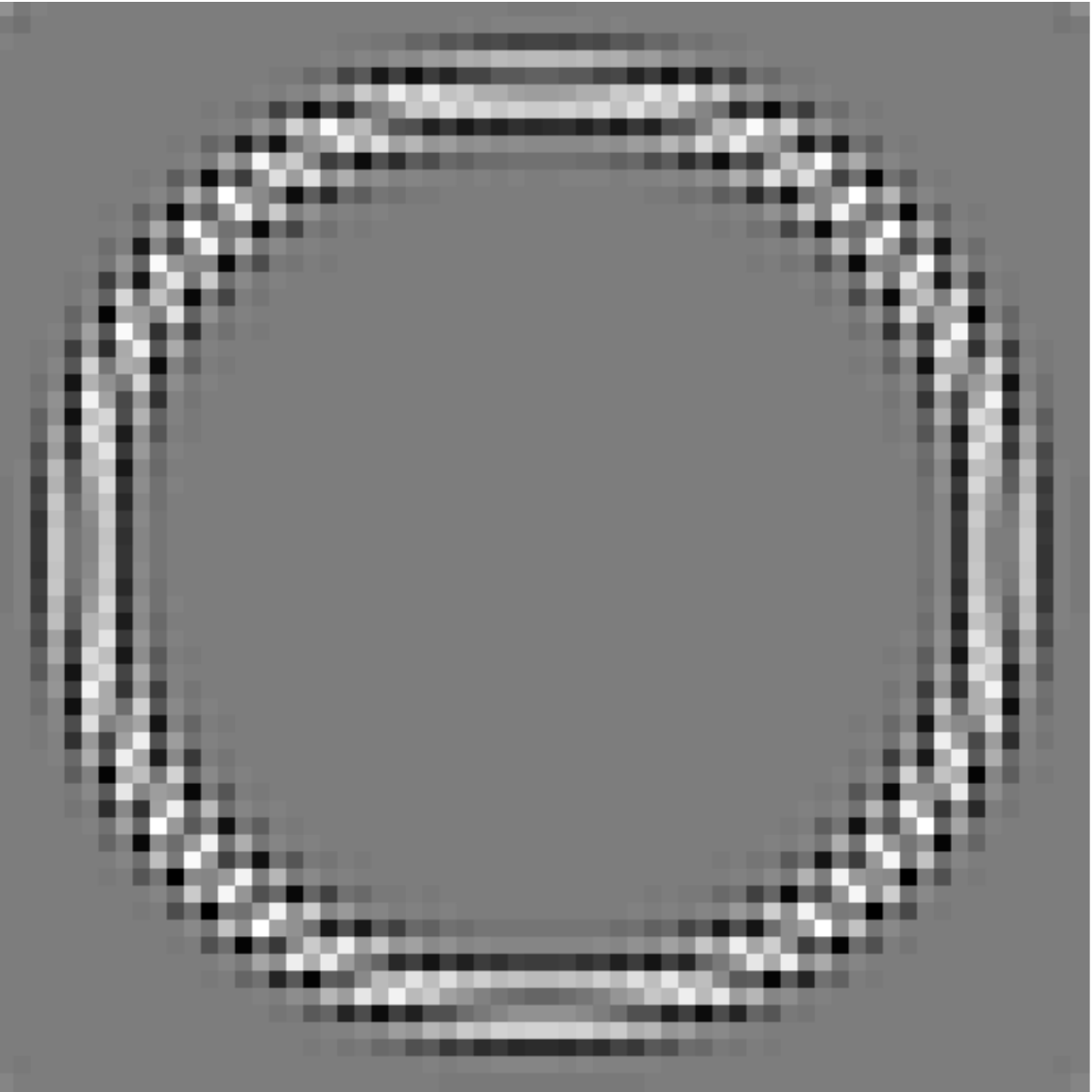}\\
&
\includegraphics[width=2.3cm]{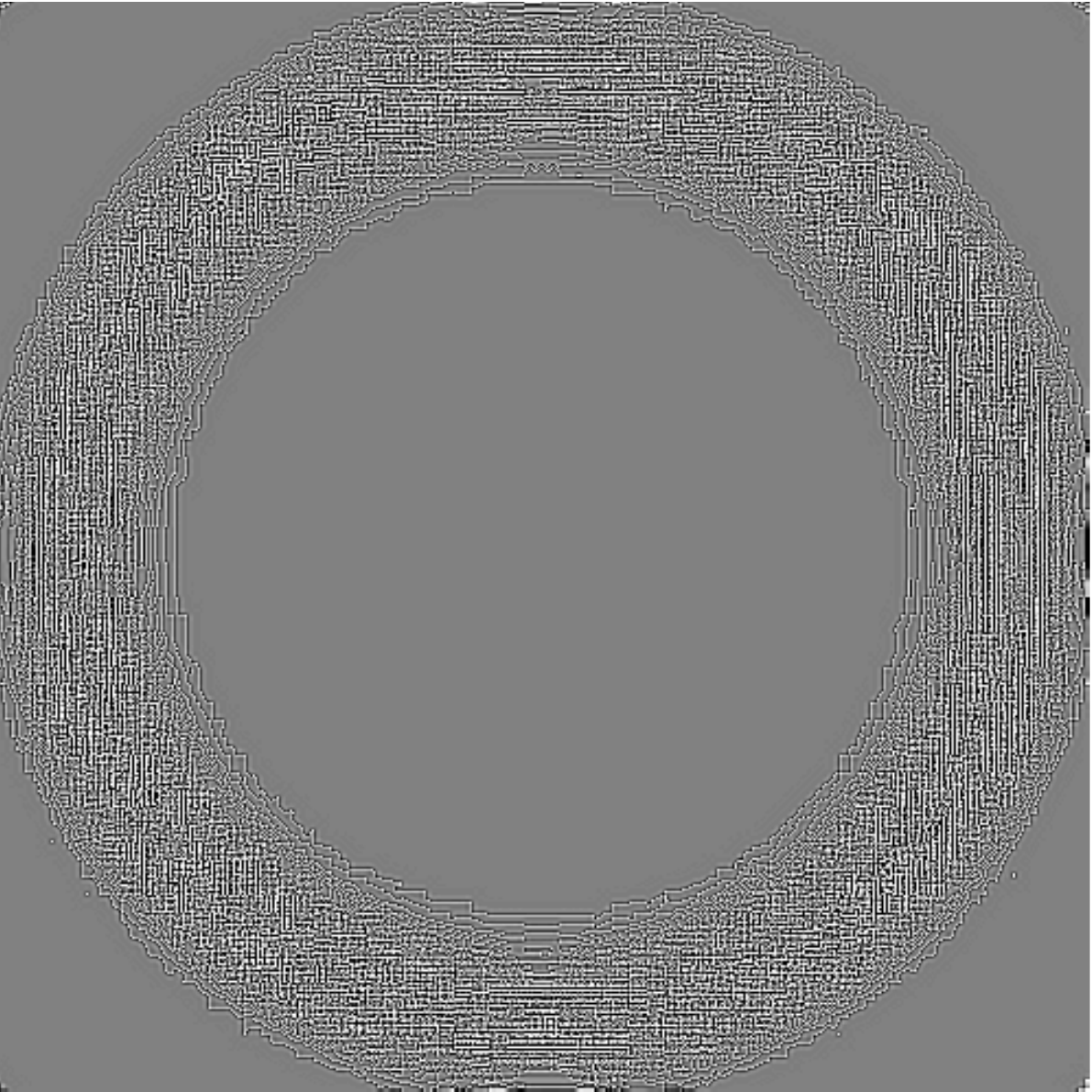}&
\includegraphics[width=2.3cm]{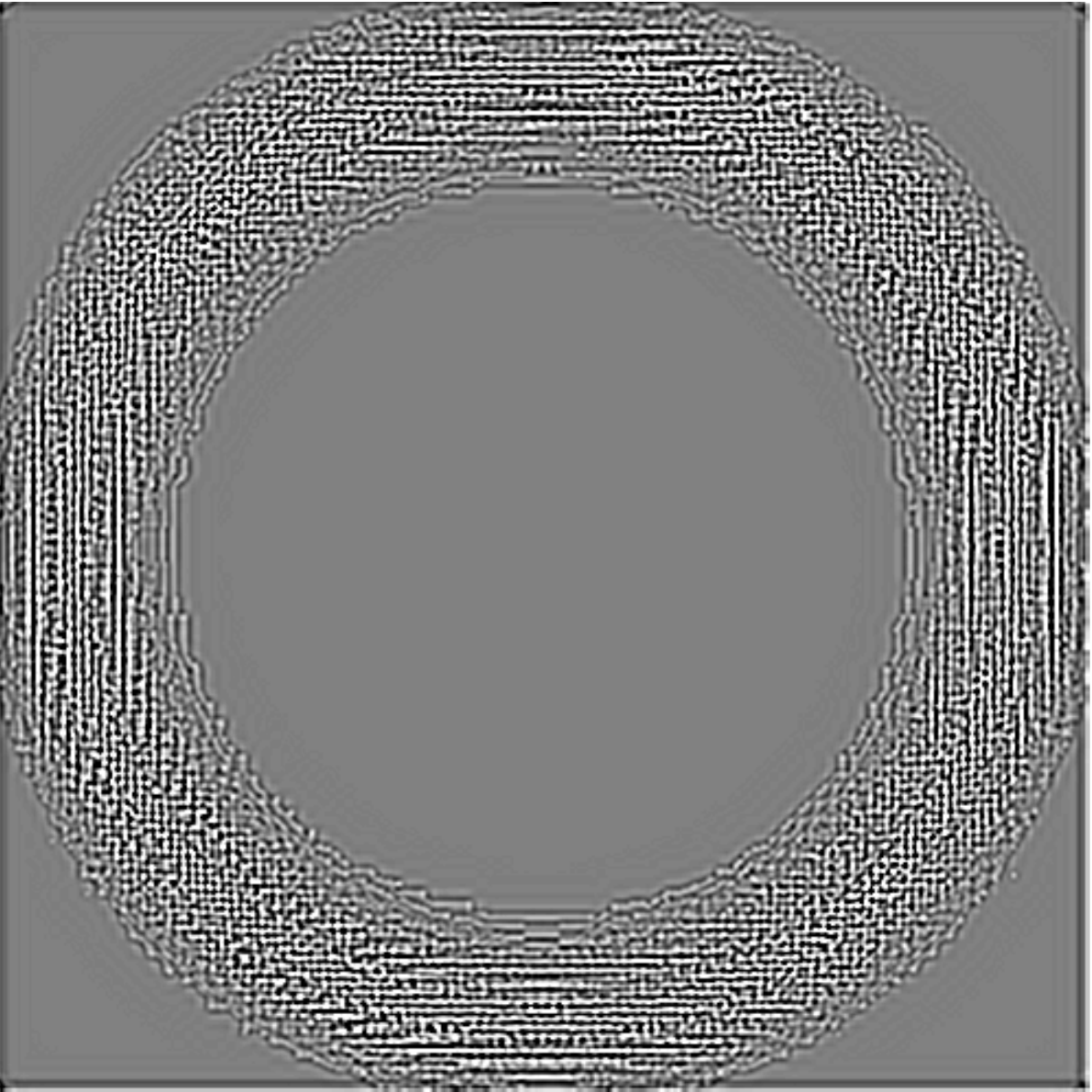}&
\includegraphics[width=2.3cm]{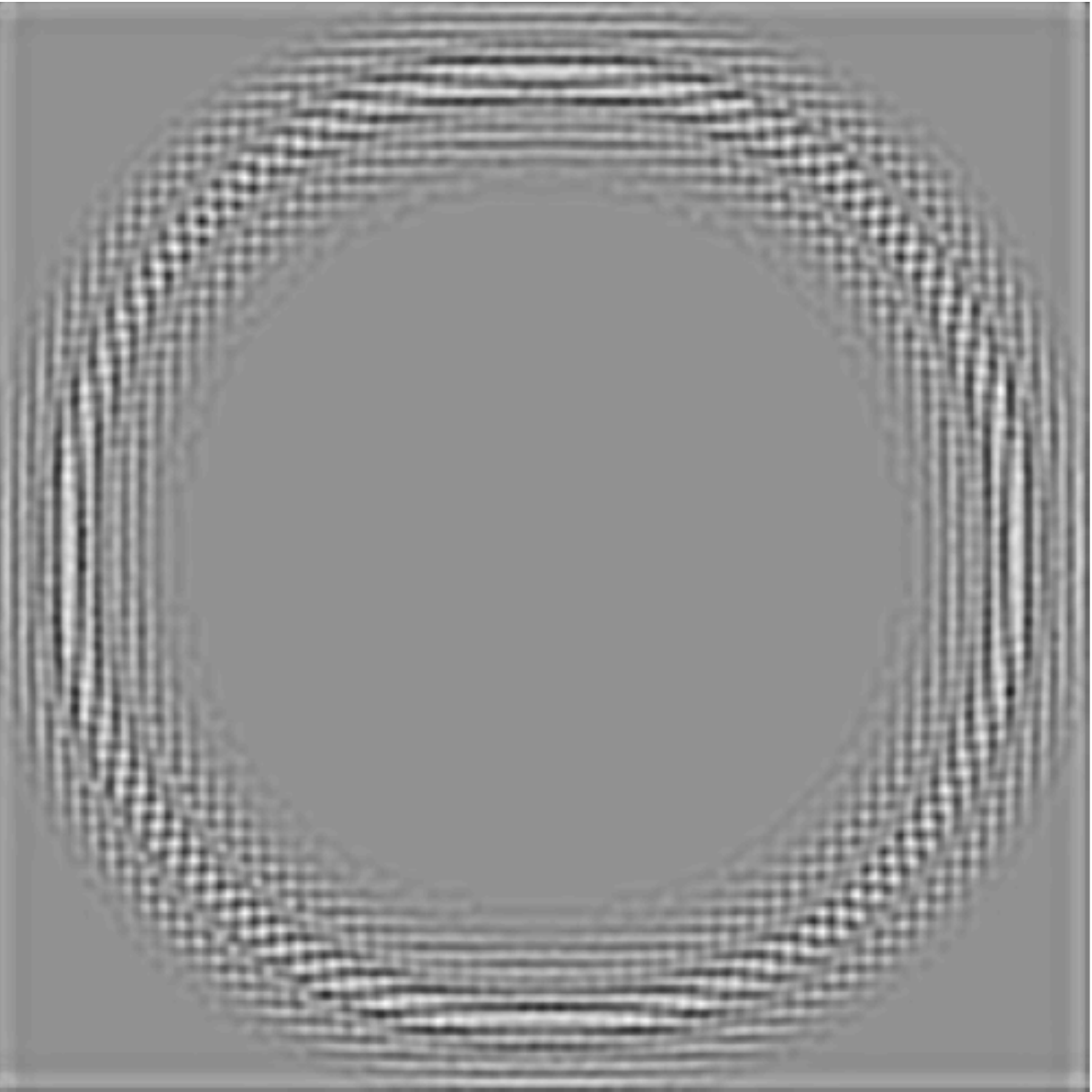}&
\includegraphics[width=2.3cm]{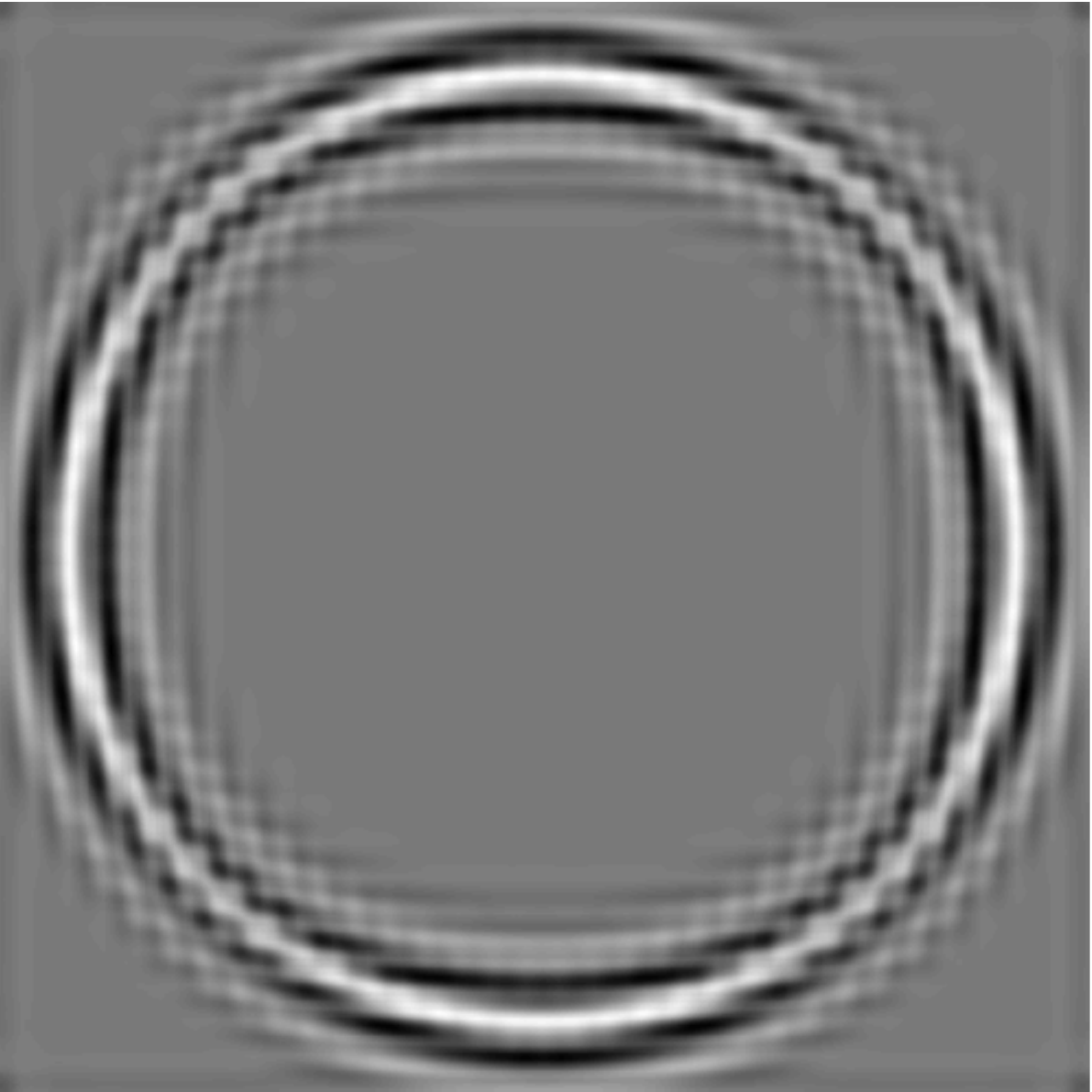}\\
&
\includegraphics[width=2.3cm]{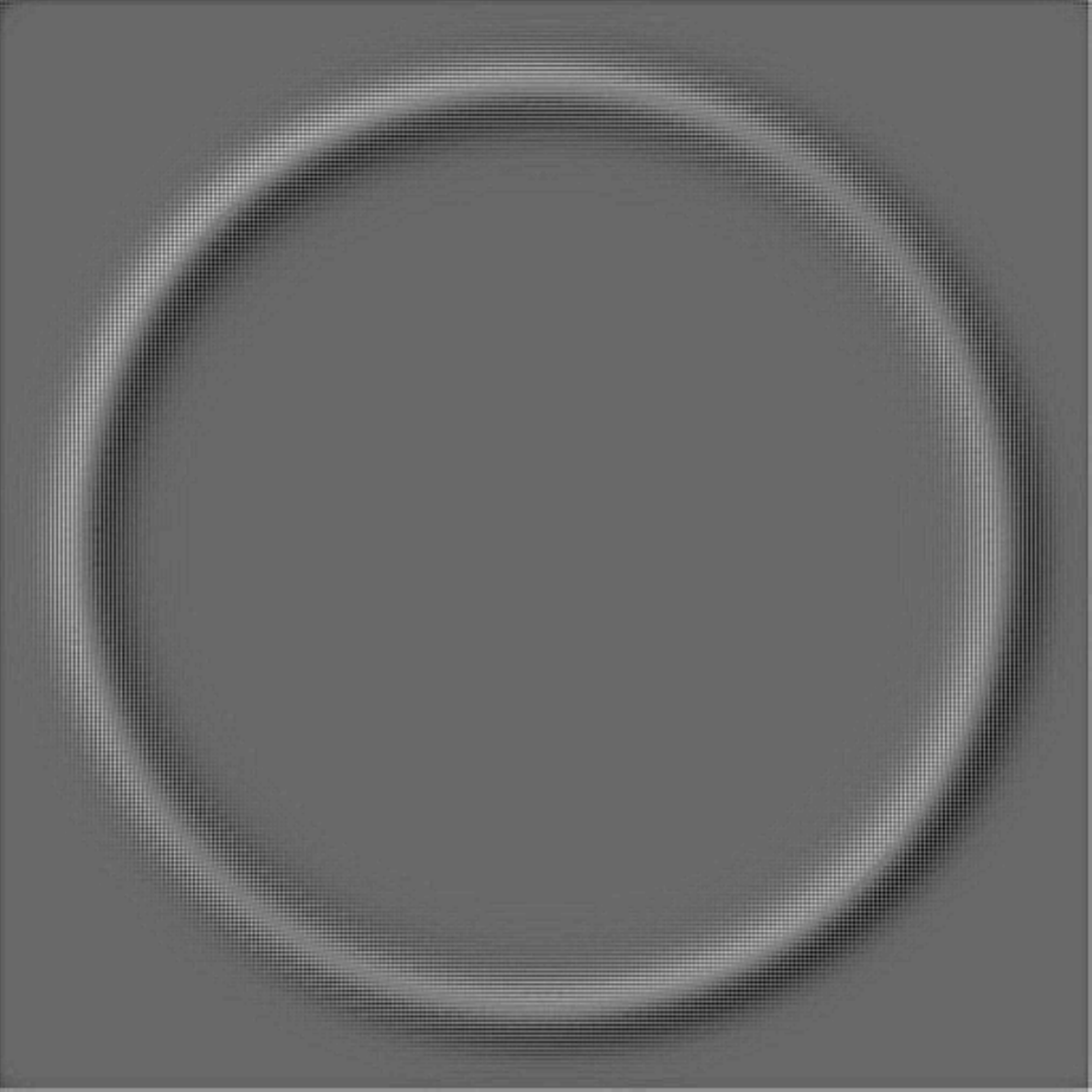}&
\includegraphics[width=2.3cm]{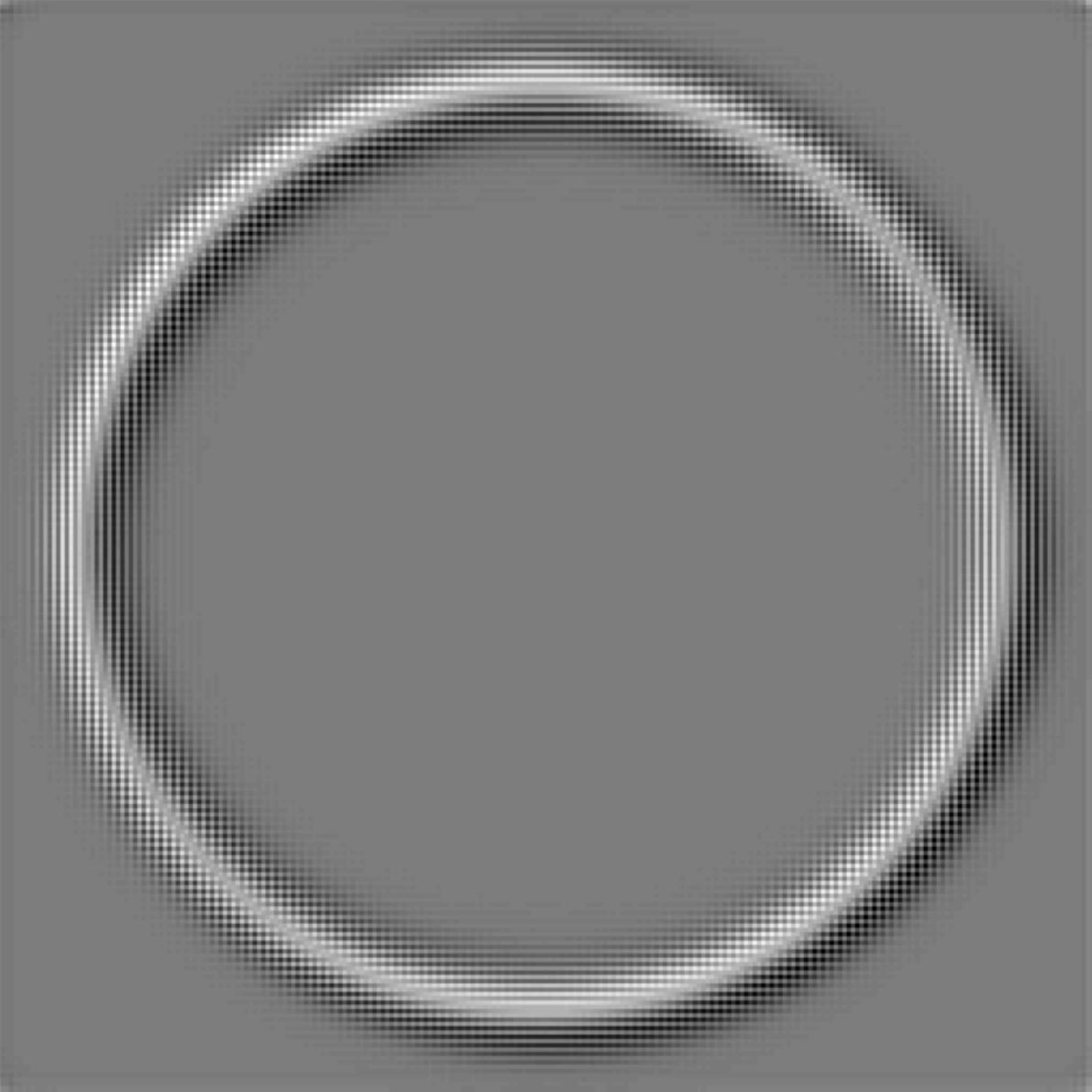}&
\includegraphics[width=2.3cm]{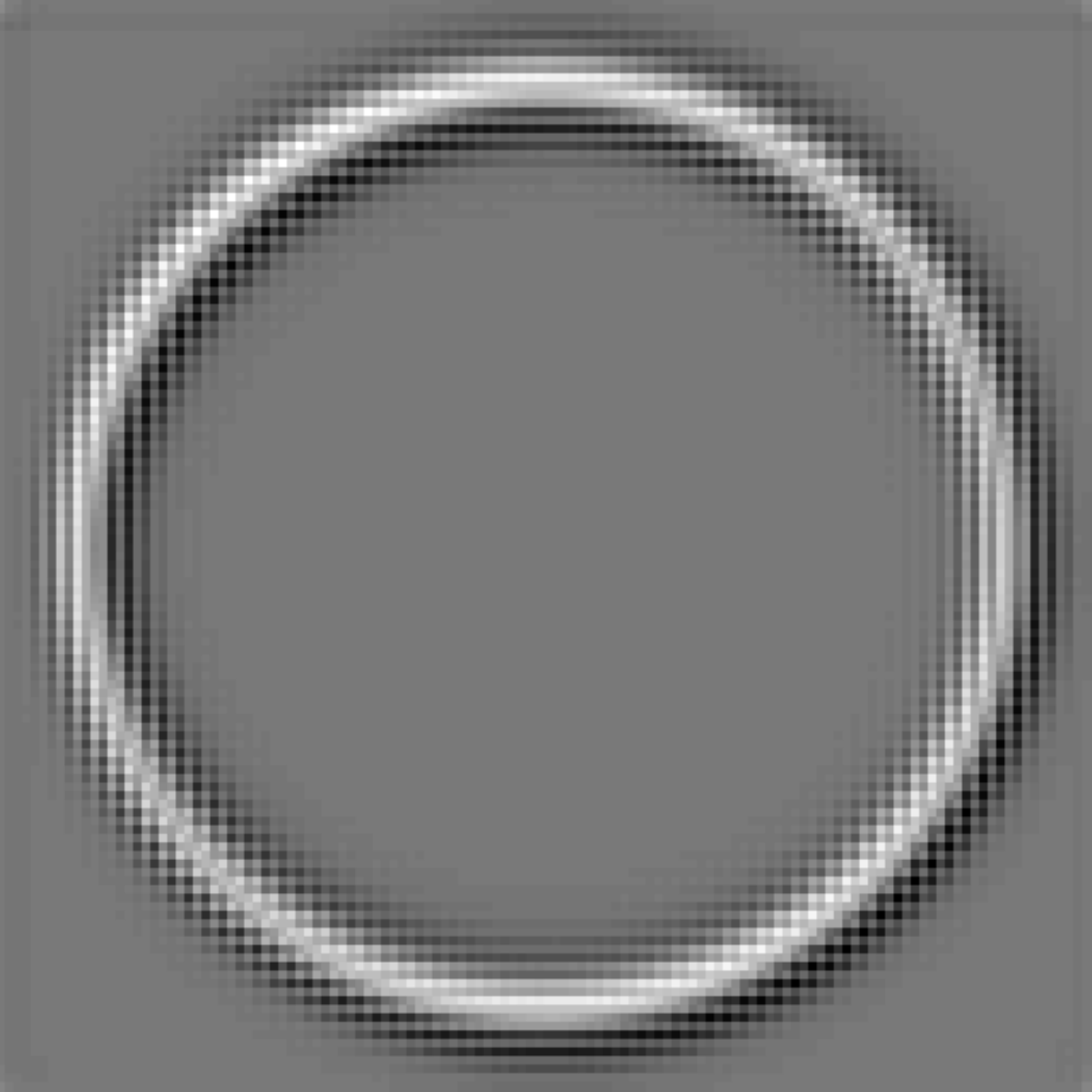}&
\includegraphics[width=2.3cm]{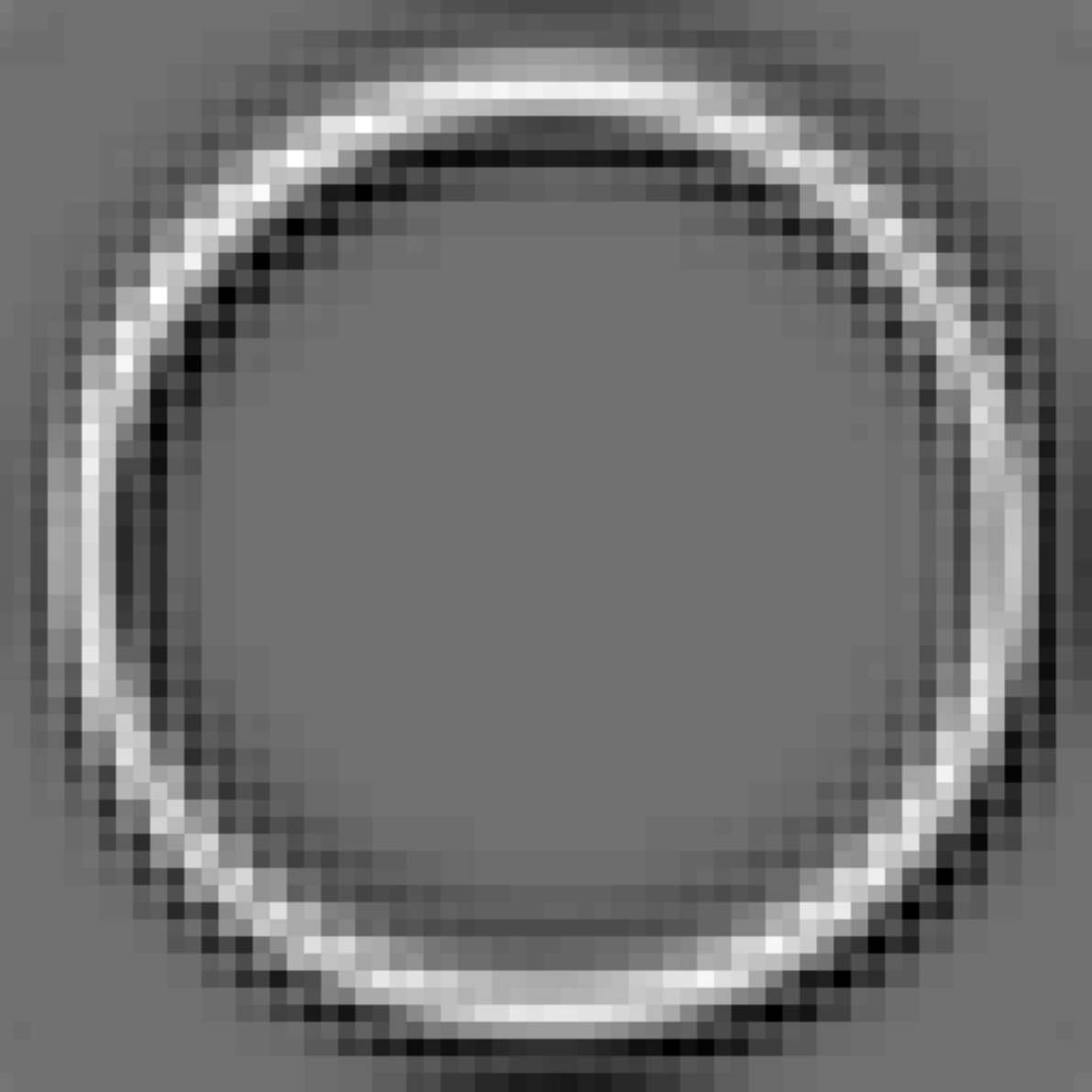}
\end{tabular}
\end{center}
\caption{Original image and its reconstructed images with only detail components at decomposition levels $J=1,\dots, 4$.
Top row: DWT, middle row: DT$\mathbb{C}$WT, bottom row: DLWT.}
\label{decom_recon.fig}
\end{figure*}

\subsection{Edge detection}
We now show examples of edge detection with a comparison to two conventional methods, namely, the standard DWT and the dual-tree complex wavelet transform (DT$\mathbb{C}$WT) \cite{dualtree}.
Figure \ref{decom_recon.fig} represents reconstructed images with only detail components at each decomposition level $J=1,\dots, 4$.
More precisely, for example $J=4$, we run Algorithm \ref{alg1} to obtain the coarse component $\{c_{j-4}[\vt]\}_{\bm{t}\in\ZZ^2}$ and the detail components
 $\{d_{k,j-J}[\vt]\}_{\bm{t}\in\ZZ^2,k\in D, 1\leq J\leq 4}$.
Then, we set all of the coarse components $\{c_{j-4}[\vt]\}_{\bm{t}\in\ZZ^2}$ to zero.
Finally, we apply Algorithm \ref{alg2} with these coarse and detail components, which obtains the image represented with only edge components.

In the bottom row of Figure \ref{decom_recon.fig}, we observe that the edges of an image of a circle represented by the proposed DLWT are fully detectable at every decomposition level $J$.
In contrast, we see that, in the case of the conventional methods, i.e., the DWT and DT$\mathbb{C}$WT, very few edges are detected at high resolution.
For lower resolution, e.g., $J=4$, we see that edges seem to be well detected. However, it is an insufficient amount compared to the case considered herein.
We consider that these results support the superiority of our edge detection over conventional methods.

\begin{figure*}[t]
\scriptsize
\begin{center}
\begin{tabular}{ccc}
\includegraphics[width=3.5cm]{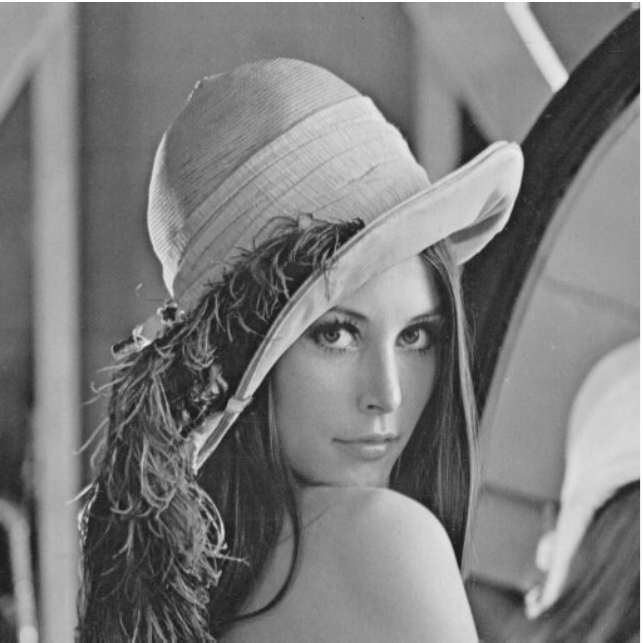}&
\includegraphics[width=3.5cm]{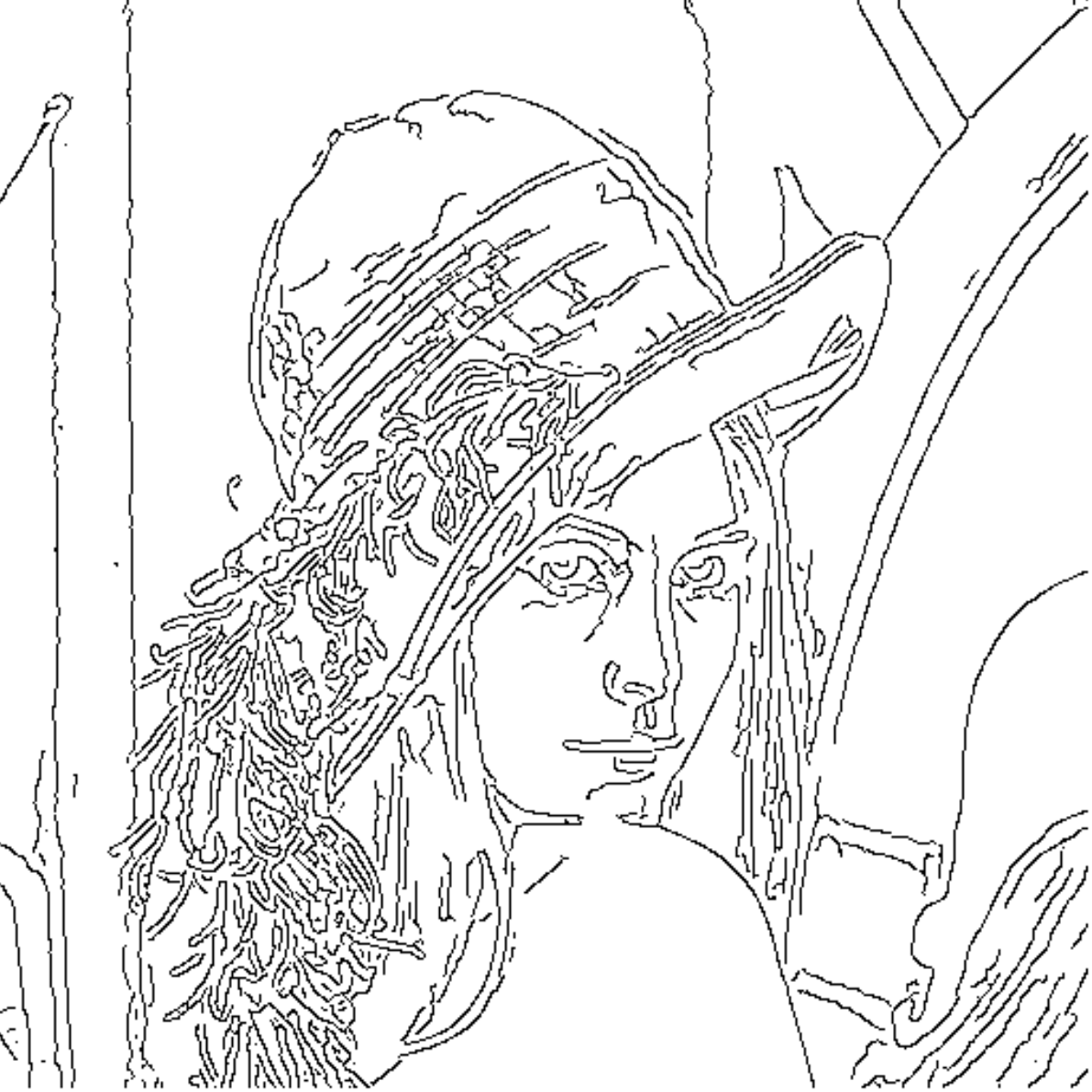}&
\includegraphics[width=3.5cm]{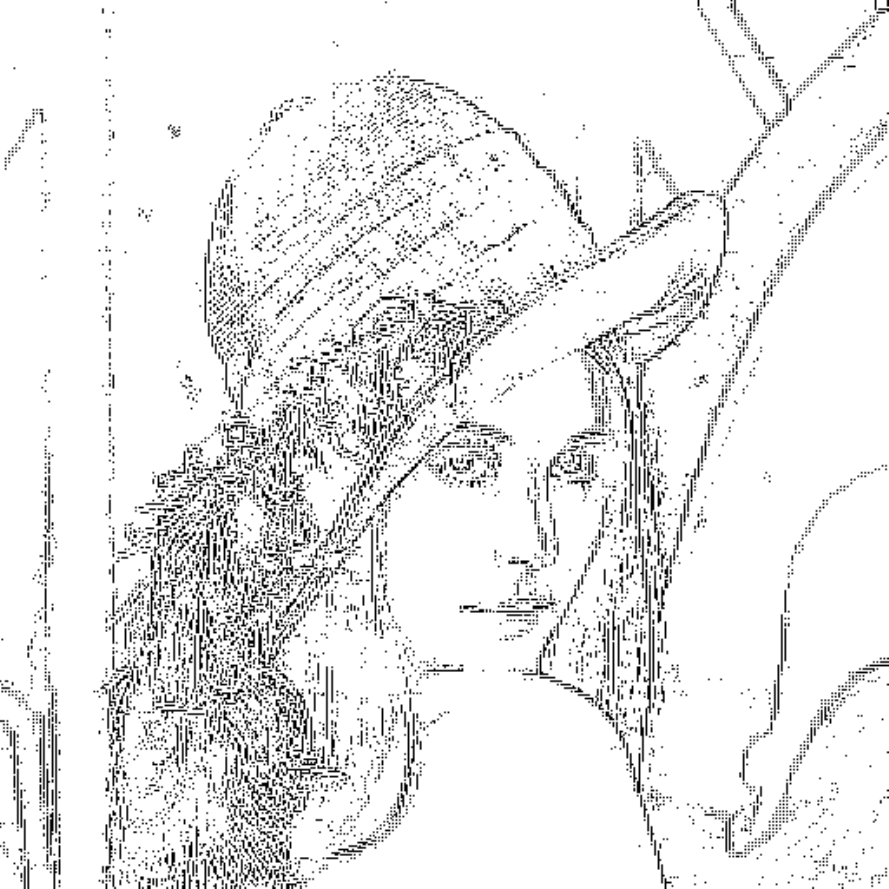}\\
(a) Original & (b) Canny & (c) DWT \\
\includegraphics[width=3.5cm]{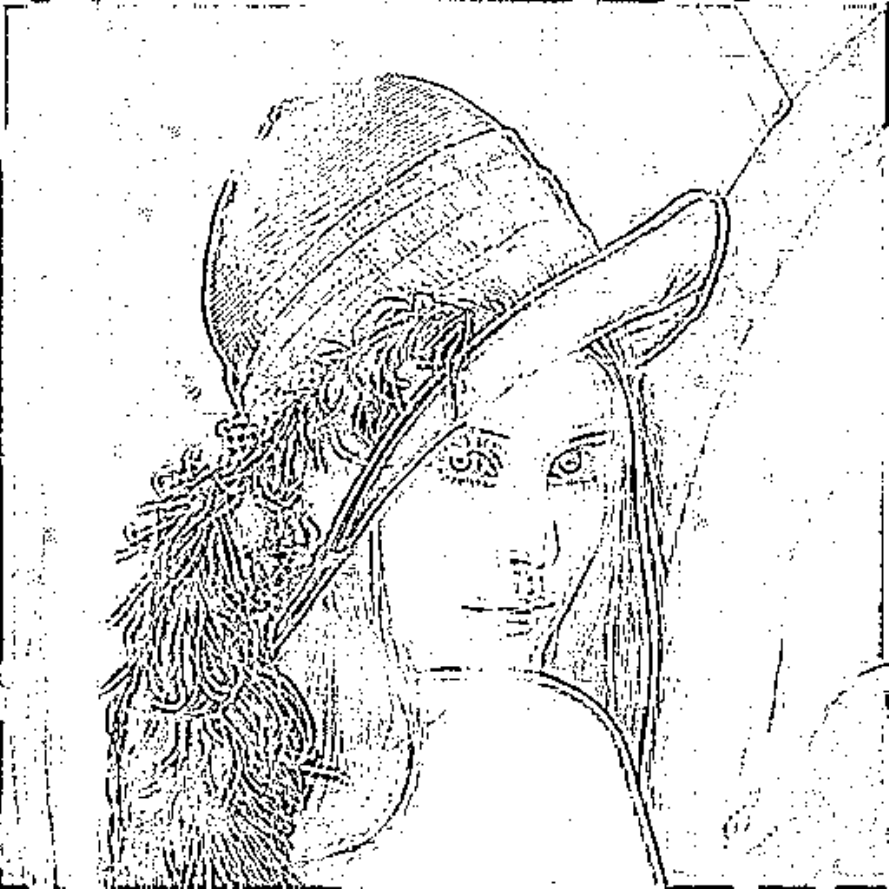}&
\includegraphics[width=3.5cm]{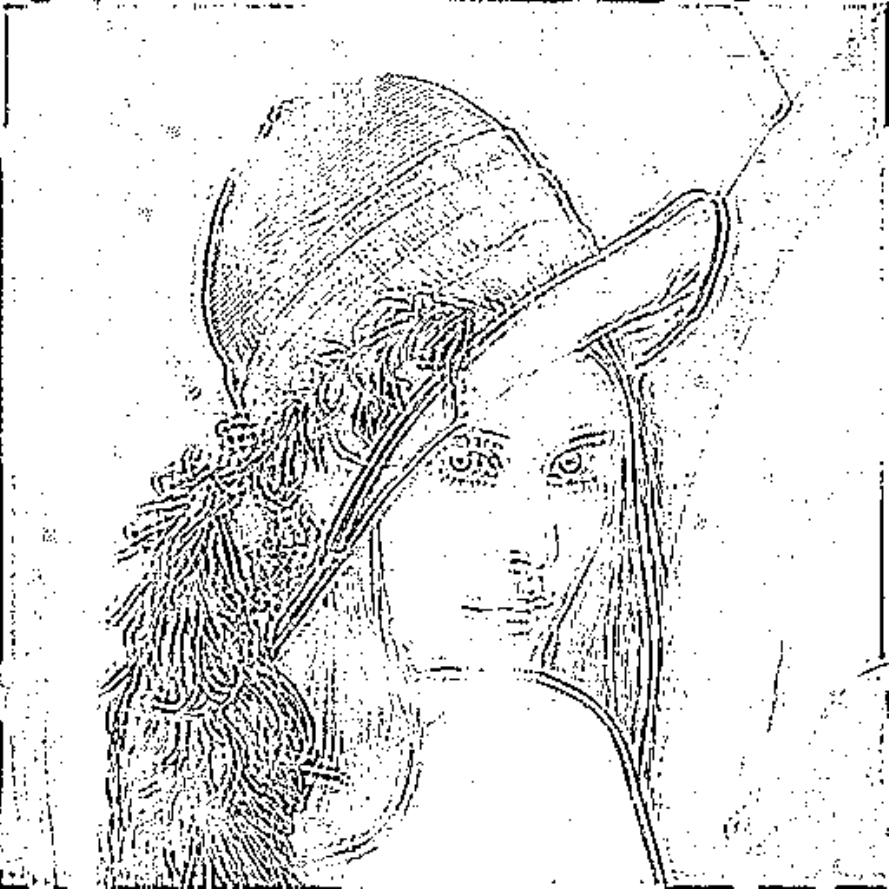}&
\includegraphics[width=3.5cm]{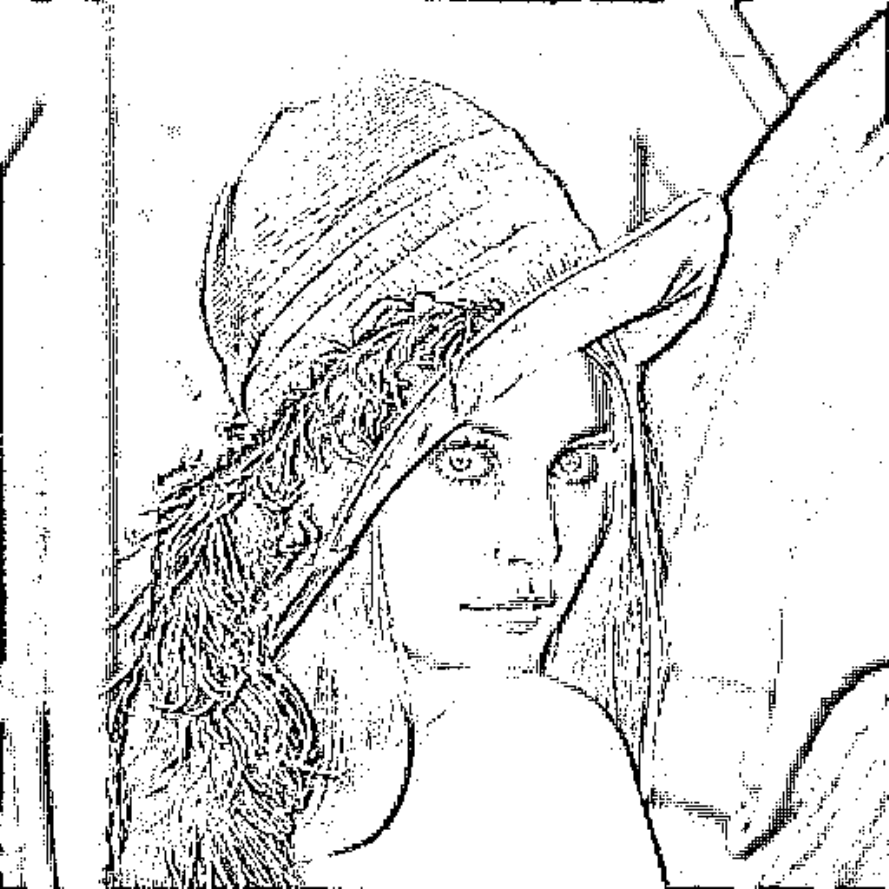}\\
(d) DT$\mathbb{C}$WT & (e) Shearlet & (f) DLWT\\
\end{tabular}
\end{center}
\caption{Results of edge detection.}
\label{recon_lena.fig}
\end{figure*}

The next example is our main result for edge detection application.
The results in Figures \ref{decom.fig} and \ref{decom_recon.fig} showed that the proposed method can offer nearly isotropic detection of edges. 
We updated the edge detection method demonstrated in Figure \ref{decom_recon.fig} and applied this method to natural images.
Two representative conventional methods of edge detection, namely, the Canny filter \cite{canny} and the shearlet approach, will also be compared.
The updated method is to binarize the edge components extracted in Figure \ref{decom_recon.fig} by thresholding in order to extract only the purer edge components.
In addition, we selected an optimal decomposition level for each image in order to maximize the edge detection performance.

The results for the updated method for with Lena image are shown in Figure \ref{recon_lena.fig}.
At a quick glance, we can see that the classical Canny edge detector is a well-balanced edge detection method, but, compared to the wavelet-based method used here, this method is insufficient for detecting detailed edges.
Since these two methods belong to completely different categories, we focus only on comparisons of four wavelet-based methods including ours.
The DT$\mathbb{C}$WT and the shearlet are better than the DWT because these two methods were originally developed to improve the directional selectivity of the DWT.
Note here that the proposed method achieves much better edge detection than these methods, which can be clearly confirmed by a glance at Figure \ref{recon_lena.fig}.
In the case of the DLWT, we use a Gaussian filter for smoothing because this filter may detect too many edge components.
Note also that this increases the processing steps compared to other conventional methods, but the computational cost remains low thanks to fast implementation by Algorithm \ref{alg1}.

\begin{figure}[t]
\scriptsize
\begin{center}
\begin{tabular}{ccc}
\includegraphics[width=3.5cm]{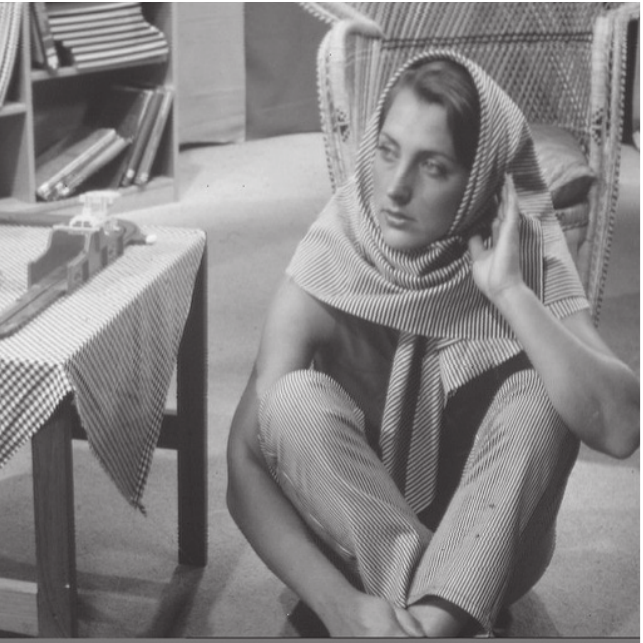}&
\includegraphics[width=3.5cm]{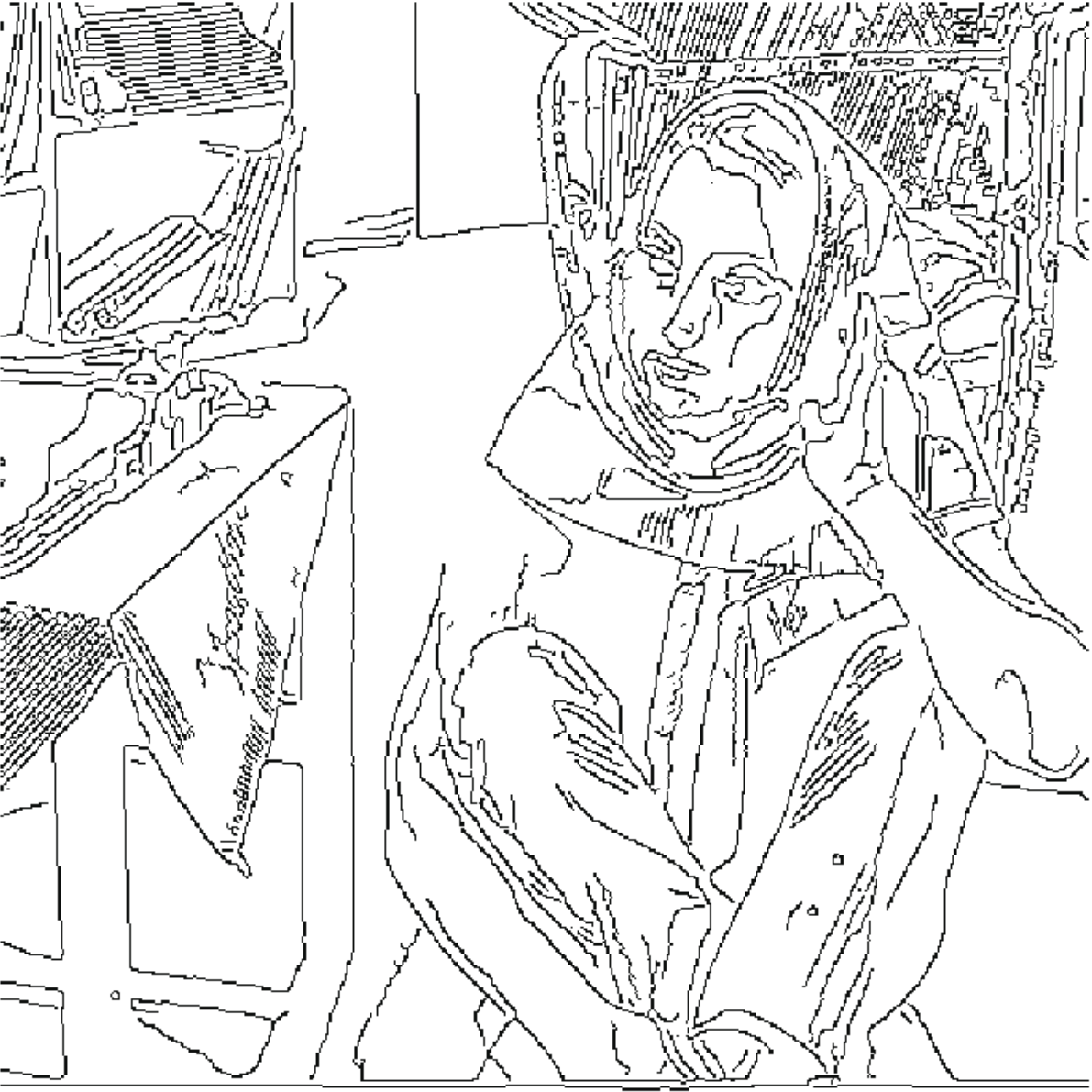}&
\includegraphics[width=3.5cm]{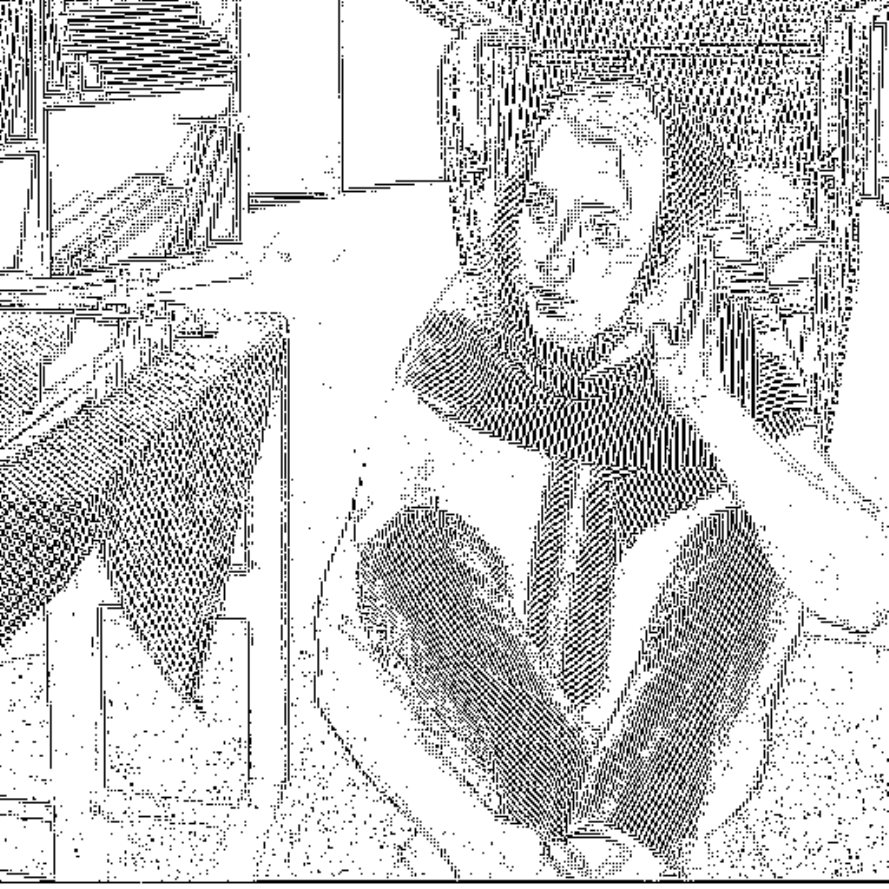}\\
(a) Original & (b) Canny & (c) DWT \\
\includegraphics[width=3.5cm]{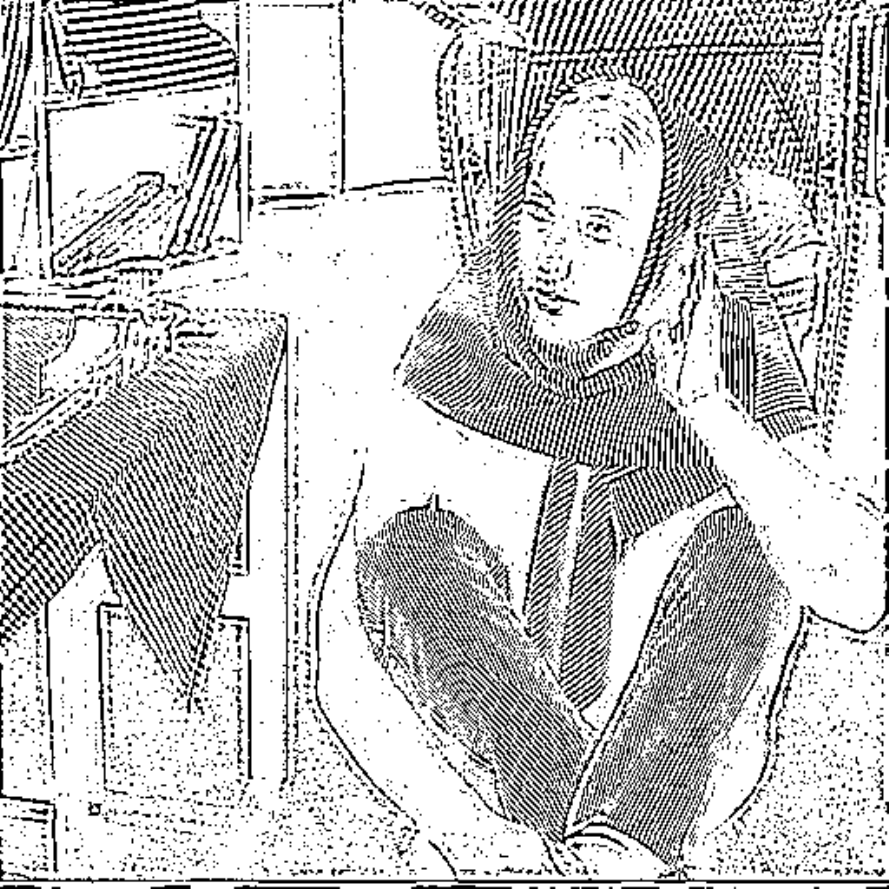}&
\includegraphics[width=3.5cm]{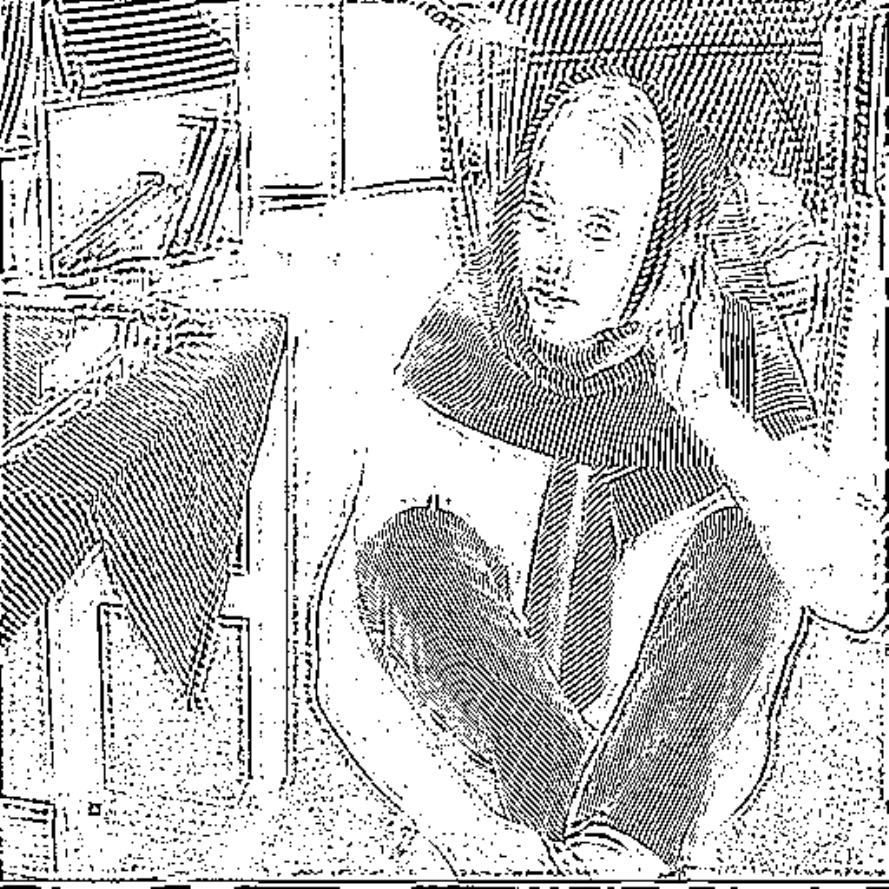}&
\includegraphics[width=3.5cm]{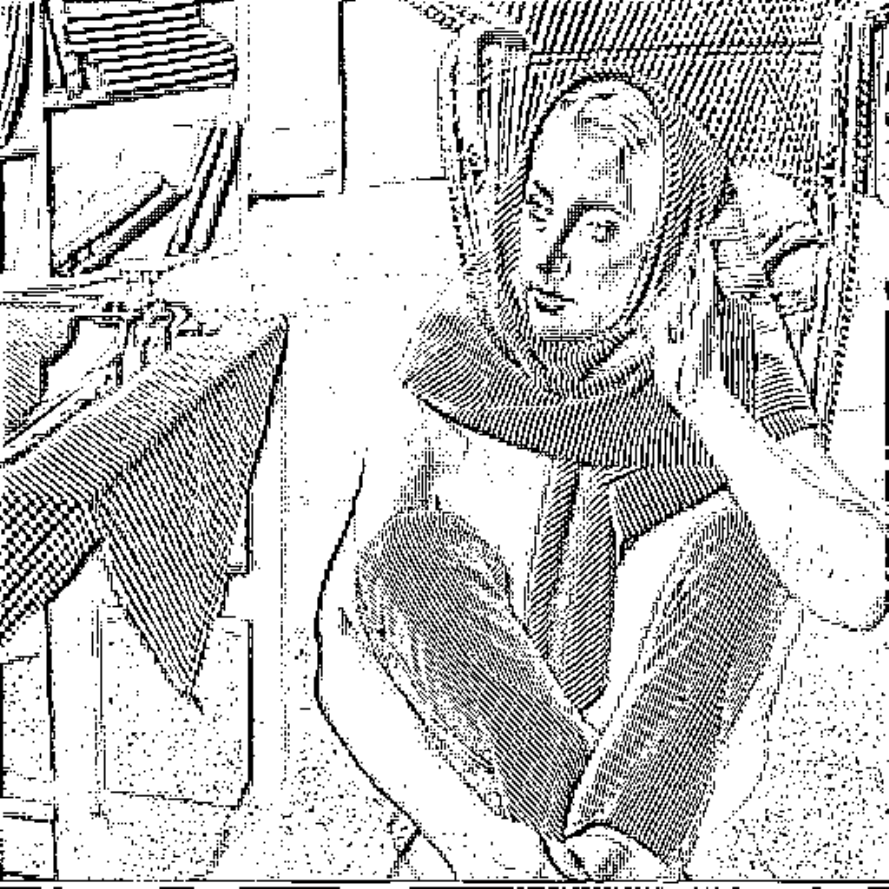}\\
(d) DT$\mathbb{C}$WT & (e) Shearlet & (f) DLWT\\
\end{tabular}
\end{center}
\caption{Results of edge detection.}
\label{recon_barbara.fig}
\end{figure}

Similar results are observed for the Barbara image shown in Figure \ref{recon_barbara.fig}.
For conventional wavelet-based methods, edges that contain high-frequency components, such as the clothes of Barbara, are well detected, especially for the cases of the DT$\mathbb{C}$WT and the shearlet.
However, some local characteristics, such as the face of Barbara, are not well represented.
This is in stark contrast to the case of the DLWT, in which both clothes and the face are well detected, indicating the huge advantage of the DLWT.
Summarizing the results in this section, it is clear that the edge detection capability of the DLWT is quite high.

\section{Concluding remarks}
In this paper, we proposed a new directional wavelet transform for directional analysis of an image.
The proposed DLWT significantly improves the directional selectivity of the classical DWT, providing nearly isotropic signal decomposition into 12 directions.
Although the DLWT is a redundant transform, its computational cost is not high because the redundancy of the DLWT is limited and a fast in-place computation algorithm based on a modified lifting scheme is available.
Therefore, the DLWT is more efficient and superior to several conventional methods in terms of the trade-off between computational cost and directional selectivity.
In addition, the DLWT inherits the nature of the lifting scheme, which guarantees the invertibility of the transform and provides a high degree of freedom in designing filters for the transform.

The results of image decomposition showed the excellent directional selectivity of the proposed method.
We also proposed a simple edge detection method that takes into account the features of the DLWT.
Numerical experiments on edge detection involving a comparison with several conventional edge detection methods demonstrated the advantages of the proposed method in terms of capturing both global and local edge structures well.

In the present paper, we dealt with an application intended for uniform extraction of edges, but it is also possible to add anisotropy to the transform by using lifting operators with different properties for each direction in the design of the filters in the DLWT. 
In the near future, it will be necessary to study the inclusion of such possibilities and more general complex data analysis by using the correlation of data in the spatial domain.

\section*{Acknowledgment}
The present work was supported in part by JSPS KAKENHI (Grant Number 21K11945 and 21K11972).

\end{document}